\documentclass[smus]{snow2e}
\usepackage{graphics}
\input{epsf}

\def\ar#1#2#3{     {\it Ann. Rev. Nucl. and Part. Sci. }{\bf #1}, #2 (19#3)}

\def\np#1#2#3{           {\it Nucl. Phys. }{\bf #1}, #2 (19#3)}
\def\pl#1#2#3{           {\it Phys. Lett. }{\bf #1}, #2 (19#3)}
\def\pr#1#2#3{           {\it Phys. Rev. }{\bf #1}, #2 (19#3)}

\def\prl#1#2#3{          {\it Phys. Rev. Lett. }{\bf #1}, #2 (19#3)}   

\def\rmp#1#2#3{          {\it Rev. Mod. Phys. }{\bf #1}, #2 (19#3)}

\def\zp#1#2#3{           {\it Zeit. fur Physik }{\bf #1}, #2 (19#3)}
	\def\etc{\hbox{\it etc.}}
\def\eg{\hbox{\it e.g.}}	
\def\etal{\hbox{\it et al.}}

\def\etal{\hbox{\it et al.}}

\def \to {\rightarrow}

\def\dofig#1{\vskip.2in\centerline{\epsfbox{#1}}}
\input rotate
\def \et {E_{T}}

\def \ETmiss {{E}_{T}^{miss}}
\def\simge
{\mathrel{\rlap{\raise 0.53ex \hbox{$>$}}{\lower 0.53ex \hbox{$\sim$}}}}

\def\simle
{\mathrel{\rlap{\raise 0.4ex \hbox{$<$}}{\lower 0.72ex \hbox{$\sim$}}}}

\def \gluino {\tilde{g}} 
\def \squark {\tilde{q}}

\def  \met {\not\!\!\et }

\def  \abseta {\mid\eta\mid}

\begin{document}

\title{
\begin{flushright}
{\large{LBL-38997}}\\
{\large{FERMILAB-Conf-96/432}}\\
\end{flushright}
High Transverse Momentum Physics at the Large Hadron Collider
\thanks
{To appear in the Proceedings of the 1996 DPF/DPB Summer Study on
New Directions for High Energy Physics (Snowmass 96).}
}

\author{The U.S.~ATLAS and U.S.~CMS Collaborations\\
{\textit{edited by}}\\
Ian~Hinchliffe\\ {\textit{E.O. Lawrence Berkeley National 
Laboratory, Berkeley, CA 94720}}\\
John~Womersley\\ {\textit{Fermi National Accelerator
Laboratory, Batavia, IL 60510}}\\
}
\setlength{\titleblockheight}{6cm}
\maketitle

%% Get rid of page numbering on first page
\thispagestyle{empty}
\pagestyle{plain}

%%%%% sample figure using Snowmass graphics
%\begin{figure}[t]
%\leavevmode
%\begin{center}
%\vspace{2.5cm}
%\resizebox{!}{8cm}{%
%\includegraphics{motivate_eta_e_cut.ps}}
%\end{center}
%\vspace{-2.5cm}
%\caption{Reconstructed lepton pseudorapidity distributions for signal
%and $W+$jets processes.}
%\label{fig:eta_e}
%\end{figure}
%%%

\begin{abstract}
This note summarizes the various physics studies done for the LHC. It concentrates on the processes involving the 
production of high mass states. Results are drawn from simulations performed by the CMS and ATLAS collaborations.
The ability of the LHC to provide insight into the mechanism of electroweak symmetry breaking is exemplified.
\end{abstract}
\section{Introduction and Motivation}

This document is intended to summarize the potential of the Large Hadron
Collider (LHC) for high transverse momentum physics and explain the reasons why
it is a crucial next step in our understanding of the behavior of nature.
It is the physics potential of the LHC that motivates US participation;
not a desire to build detector components, a need for projects for students or
postdocs, or a requirement for a future program to retain university funding
(though these elements may be important).  We believe this physics potential enormous and
that, among currently approved projects, the LHC  is unique in that it is the only one 
that has sufficient energy and luminosity to probe in detail the energy scale 
relevant to electroweak symmetry breaking.

We outline the many physics processes that have been studied as part of the design processes for the 
ATLAS\cite{atlas} and CMS\cite{cms} detectors. Examples are selected from the 
large amount of detailed work carried out for and since the technical proposals.

\subsection{The Standard Model}

The Standard Model (SM) is a very successful description of the interactions of
the components of matter at the smallest scales ($\simle 10^{-18}\,$m) and
highest energies ($\sim 200\,$GeV) accessible to current experiments. 
It is a quantum field theory which describes the interaction of spin-$1\over 2$,
point-like fermions, whose interactions are mediated by spin-1 gauge bosons.
The bosons are a consequence of local gauge invariance applied to the fermion
fields and are a manifestation of the symmetry group of the theory, which for
the SM is $SU(3) \times SU(2) \times U(1)$.  

The fundamental fermions are leptons and quarks.  The left-handed states are
doublets under the $SU(2)$ group, while the right-handed states are singlets.
There are three generations of fermions, each generation identical except for
mass:  the origin of this structure, and the breaking of generational symmetry 
(flavor symmetry) remain a mystery.  There are three leptons with electric
charge $-1$, the electron ($e$), muon ($\mu$) and tau lepton ($\tau$),
and three electrically neutral leptons (the neutrinos $\nu_e$, $\nu_\mu$ and
$\nu_\tau$).  Similarly there are three quarks with electric charge 
$+{2\over3}$, up ($u$), charm ($c$) and top ($t$), and three with
electric charge $-{1\over3}$, down ($d$), strange ($s$) and bottom ($b$). 
The quarks are triplets under the $SU(3)$ group and thus carry an additional
``charge,'' referred to as color.  There is mixing between the three
generations of quarks, which in the SM is parameterized by the 
Cabibbo-Kobayashi-Maskawa (CKM)\cite{ckm} matrix but not explained.

In the SM the $SU(2)\times U(1)$ symmetry group (which describes the
so-called Electroweak interaction) is spontaneously broken
by the existence of a (postulated) Higgs field with non-zero expectation value.
This leads to the emergence of massive vector bosons, the $W^\pm$ and $Z$, which
mediate the weak interaction, while the photon of electromagnetism remains
massless.  One physical degree of freedom remains in the Higgs sector, which
should be manifest as a neutral scalar boson $H^0$, but which is presently
unobserved.  The $SU(3)$ group describes the strong interaction (quantum
chromodynamics or QCD). Eight vector gluons mediate this interaction. They
carry color charges themselves, and are thus self-interacting.  This implies
that the QCD coupling $\alpha_S$ is small for large momentum
transfers but large for small momentum transfers, and leads to the confinement of
quarks inside color-neutral hadrons.  Attempting to free a quark produces a jet
of hadrons through quark-antiquark pair production and gluon bremsstrahlung. 

The basic elements of the Standard Model were proposed in the 1960's and 1970's \cite{standard-model}.
Increasing experimental evidence of the correctness of the model accumulated through 1970's and 1980's:
\begin{itemize}
\item SLAC deep inelastic scattering experiments showed the existence of
point-like scattering centers inside nucleons, later identified with quarks \cite{dis}
\item observation of the $c$ and $b$ quarks \cite{bandc}
\item observation of neutral weak currents ($Z$ exchange)\cite{neutral-currents}
\item observation of jet structure and three-jet final states (gluon radiation)
in $e^+ e^-$ and hadron-hadron collisions\cite{3jets}
\item direct observation of the $W$ and $Z$ at the CERN SPS collider \cite{wz}
\end{itemize}

Following these discoveries, an era of consolidation has been entered.
Ever more precise experiments have been carried out at LEP and SLC which have 
provided verification of the couplings of quarks and leptons to the gauge
bosons at the level of 1-loop radiative corrections ($\sim {\cal O}(10^{-3})$).
The top quark was discovered at Fermilab in 1995, with a very large mass
($\sim 175\,$GeV).\cite{topdisc} 

Only two particles from the Standard Model have yet to be observed;
$\nu_\tau$ and the Higgs boson. Of these the latter is more important
as it holds the key to the generation of $W$, $Z$, quark and lepton masses.
Some of the SM parameters, particularly those of the CKM
matrix are not well determined. Experiments over the next
few years involving CP violation in the K\cite{ktev} and B systems\cite{bfactory} should determine these
parameters or demonstrate the SM cannot adequately explain CP violation.
There are some indications that the SM may be incomplete or inadequate in that there are
a very few experimental observations that it cannot accommodate such as 
%the 
%unexpectedly small decay 
%width of $Z\to b\overline{b}$ \cite{ztobbar}and 
the possibility that neutrino
oscillations occur\cite{neutrino-troubles}.

\subsection{Beyond the Standard Model}

The success of the standard model\cite{standard-model}
 of strong (QCD), weak and electromagnetic
interactions has drawn increased attention to its limitations. In its simplest
version, the model has 19 parameters \cite{cahn96}, the three coupling constants of 
the gauge theory $SU(3)\times SU(2)\times U(1)$, three lepton and six quark
masses, the mass of the $Z$ boson which sets the scale of weak interactions,
the four parameters which describe the rotation from the weak to the mass
eigenstates of the charge -1/3 quarks (CKM matrix). All of these parameters are determined
with varying errors. Of the two remaining,  one, a CP violating parameter
associated with  the strong interactions, must be very small. The last
parameter is associated with the mechanism responsible for the breakdown
the electroweak $SU(2)\times U(1)$ to $U(1)_{em}$. This 
can be taken as the mass of the, as yet undiscovered, Higgs boson.
The couplings of the Higgs boson are determined once its mass is given.

The gauge theory part of the SM has been well tested; but there is no
direct evidence either for or against the simple Higgs mechanism for electroweak
symmetry breaking.  All masses are tied to the mass scale of the Higgs sector.
Within the model we have no guidance on the expected mass of the Higgs boson.
The current experimental lower bound is 65 GeV. As its mass increases, the self couplings
and the couplings to the $W$ and $Z$ bosons grow\cite{Lee-quigg}. This feature has a very important consequence.
Either the Higgs boson must have a mass less than about 800 GeV or the 
dynamics of WW and ZZ interactions
with center of mass energies of order 1 TeV
will reveal new structure. It is this simple argument that sets the energy scale
that must be reached to guarantee that an experiment will be able to provide
information on the nature of electroweak symmetry breaking. 

The presence of a single elementary scalar boson is distasteful to many theorists.
If the theory is part of some more fundamental theory, which has some other larger
mass scale (such as the scale of grand unification or the Planck scale), there is
a serious ``fine tuning'' or naturalness problem. 
Radiative corrections to the Higgs boson mass result in
a value that is driven to the larger scale unless some delicate cancellation is 
engineered ($m_0^2-m_1^2\sim M_W^2$ where $m_0$ and $m_1$ are order $10^{15}$ GeV or larger).
There are two ways out of this problem which involve new physics on the scale of 1 TeV.
New strong dynamics could enter that provide the scale of $m_W$ or new particles could appear
so that the larger scale is still possible, but the divergences are cancelled on a much smaller 
scale. In any of the options, standard model, new dynamics or cancellations, the energy scale is the same;
something must be discovered on the TeV scale.

Supersymmetry is an appealing concept for which there is, at present, no experimental
evidence\cite{susy}. It offers the only presently known mechanism for incorporating gravity into
the quantum theory of particle interactions and provides an elegant cancellation
mechanism for the divergences provided that
at the electroweak scale the theory is supersymmetric.
The successes if the Standard Model
(such as precision
electroweak predictions) are retained, while avoiding any fine tuning of 
the Higgs mass. Some supersymmetric models allow for the unification of gauge couplings at a high scale and a
consequent reduction of the number of arbitrary parameters.
Supersymmetric models postulate the existence of superpartners for all
the presently observed particles: bosonic superpartners of fermions
(squarks $\squark$ and sleptons $\tilde \ell$), and
fermionic superpartners of bosons (gluinos $\gluino$ and gauginos 
$\tilde\chi^0_i$, $\tilde\chi^\pm_i$). 
There are also multiple Higgs bosons: $h$, $H$, $A$ and $H^\pm$.  
There is thus a large spectrum of presently unobserved
particles, whose exact masses, couplings and decay chains 
are calculable in the theory given certain parameters.  Unfortunately these
parameters are unknown.  Nonetheless, if supersymmetry is to have anything to do
with electroweak symmetry breaking, the masses should be in the region 100~GeV -- 1~TeV. 

An example of the strong coupling 
this scenario is ``technicolor'' or models based on dynamical symmetry breaking\cite{technicolor}.
Again, if the dynamics is to have anything to 
do with Electroweak Symmetry breaking we would expect
new states in the region 100~GeV -- 1~TeV; most models predict a large spectrum.
An elegant implementation of this appealing idea is lacking.
However, all models predict structure in the $WW$ scattering amplitude 
at around 1 TeV center of mass energy. 

There are also other possibilities for new physics that are not necessarily related to the scale of electroweak symmetry breaking.
There could be new neutral or charged gauge bosons with mass larger than the $Z$ 
and $W$; there could be
new quarks, charged leptons or massive neutrinos; 
or quarks and leptons could turn out not to be elementary objects.
While we have no definitive expectations for the masses of these objects,
the LHC must be able to search for them over its
available energy range.

%An intriguing possibility is raised by the top quark's very large mass ($\sim
%175\,$GeV).  In the Standard Model, the gauge bosons obtain mass through the
%mechanism of EWSB, while the fermion masses are generated by {\it ad hoc} Yukawa
%couplings to the Higgs field.  For all fermions known before the top quark,
%these couplings are very small; but the top coupling is of order unity.
%Thus, what once seemed two separate issues --- the generation of gauge boson
%masses and the generation of quark and lepton masses --- are now seen to operate
%on the same mass scale.  This argues that the question of flavor symmetry
%breaking cannot be postponed, but is of a piece with EWSB.  This is surely true
%in any dynamical scheme, but is an appealing conclusion in general.  If this is
%the case, then the LHC is not just a window to the mechanism of electroweak
%symmetry breaking, but also to flavor symmetry breaking and the origin of
%fermion masses --- and a great opportunity awaits us.  

\subsection{Accelerator Facilities}

High energy physics is explored experimentally by accelerating and colliding
beams of quarks and leptons.  Electrons (and/or positrons) and protons (and/or
antiprotons) are used in practice.  It is much easier to reach high energies
using protons, but each of the constituent quarks and gluons carries only a
fraction of the total energy.  

The present comprehensive state of understanding the Standard Model stems in
large part from our having a wide range of facilities which explore the
interactions between the fermions at energy scales
$\sqrt{\hat s}$ of order $m_{W,Z} \sim 100\,$GeV to $m_t \sim 180\,$GeV. 
These are:
\begin{itemize}
\item The Fermilab Tevatron collider, with $p\overline p$ collisions at
$\sqrt{s}=1.8\,$TeV;
\item The CERN LEP collider, with $e^+e^-$ collisions at
$\sqrt{s}=m_Z$, increasing to about 180~GeV in LEP~2 (1996);
\item The SLAC SLC collider, with $e^+e^-$ collisions at
$\sqrt{s}=m_Z$;
\item The DESY HERA collider, which collides 30~GeV $e^\pm$ with
800~GeV protons.
\end{itemize}
While either LEP 2 or the Tevatron may be sufficiently lucky to discover
new physics in the coming decade, 
there is {\it only one} facility under construction that will really 
enable us to address interactions at energy scales 250~GeV -- 1~TeV: CERN's
Large Hadron Collider.  At present, this is our only sure window on to physics
beyond the Standard Model.

\section{The Large Hadron Collider}

\subsection{Machine parameters}

THe LHC machine is a proton-proton collider that will be installed in the 26.6
km circumference tunnel
currently used by the LEP electron-positron collider at CERN \cite{lhcbook}.
The 8.4 tesla dipole magnets each 14.2 meters long (magnetic length) 
are of the ``2 in 1'' type;
the apertures for both beams have common mechanical structure and cryostat.
These superconducting 
magnets operate at 1.9K and have an aperture of 56 mm. They will
be placed on the floor in the LEP ring after removal and storage of LEP.
The 1104 dipoles and 736 quadruples support beams of 7~TeV energy and a
circulating current of 0.54 A. 

Bunches of protons separated by 25 ns and with an RMS length of 75 mm
intersect at four points where experiments are placed. Two of these 
are high luminosity regions and house the ATLAS  and CMS detectors. Two other
regions house the ALICE detector \cite{alice}, to be used for the study of heavy ion
collisions, and LHC-B\cite{lhcb}, a detector optimised for the study of B-mesons and
B-Baryons.
The beams cross at an angle of 200$\mu$rad resulting in peak luminosity of
$10^{34}$ cm$^{-2}$ sec$^{-1}$ which has a lifetime of 
10 hours. At the peak luminosity there are an average of $\sim 20 pp$ interactions
per bunch crossing. Ultimately, the peak luminosity may increase to 
$2 \times 10^{34}$ cm$^{-2}$ sec$^{-1}$.
The machine will also be able to accelerate heavy
ions resulting in the possibility of Pb-Pb collisions at 1150 TeV in the center
of mass and luminosity up to $10^{27}$ cm$^{-2}$ sec$^{-1}$.

In the $pp$ version, which will be the focus of the rest of this article,
the LHC can be thought of as a parton-parton collider with beams of partons of
indefinite energy. The effective luminosity\cite{ehlq}
of these collisions is proportional to the $pp$ luminosity and falls rapidly with
the center of mass energy of the parton-parton system. The combination of the
higher energy and luminosity of the LHC compared to the highest energy collider
currently operating, the Tevatron, implies that the accessible energy range is
extended by approximately a factor of ten.
 
\subsection{Physics Goals}

The fundamental goal is to uncover and explore the 
physics behind electroweak symmetry breaking. This involves the following specific
challenges:
\begin{itemize}
\item
Discover or exclude the Standard Model Higgs and/or the multiple Higgses of
supersymmetry.
\item
Discover or exclude supersymmetry over the entire theoretically allowed mass
range.
\item
Discover or exclude new dynamics at the electroweak scale
\end{itemize}
The energy range opened up by the LHC gives us the opportunity to search for other,
possibly less well motivated, objects:
\begin{itemize}
\item Discover or exclude any new electroweak gauge bosons with masses below several TeV.
\item Discover or exclude any new quarks or leptons that are kinematically accessible.
\end{itemize}
Finally we have the possibility of exploiting the enormous production rates for
certain standard model particles to conduct the following studies:
\begin{itemize}
\item  The decay properties of the top quark, limits on exotic decays such as $t\to c Z$ or
$t\to b H^+$.
\item $b$-physics, particularly that of B-baryons and $B_s$ mesons.
\end{itemize}

An LHC experiment must have the ability to find the unexpected. New phenomena of whatever type
will decay into the particles of the standard model. In order to cover the lists given above
a detector must have great flexibility. The varied physics signatures for these processes 
require the ability to reconstruct and measure final states involving the following
\begin{itemize}
\item Charged leptons including the tau.
\item The electroweak gauge bosons $W,Z$ and $\gamma$.
\item Jets coming from the production at high transverse momentum of quarks and gluons.
\item Jets that have b-quarks within them.
\item (Missing transverse) Energy carried off by weakly interacting neutral particles such
as neutrinos. 
\end{itemize}
Particle ID which is required for a detailed study of b-physics, as opposed to b-tagging,
 is not part of ATLAS  or CMS.

In the discussion of physics signals that we present below, it is necessary to estimate 
production cross sections for both signal
and background  processes. These are estimated using perturbative QCD and depend on several ingredients.
Differences can arise from the structure functions that are used;
the energy ($Q^2$ scale) used in the evaluation of the QCD coupling constant and the structure functions;
and the order in QCD perturbation theory that is used in the calculation of the underlying parton process.
These issues can make comparison between different simulations of the same process difficult.
Higher order corrections are not known for all processes and in some cases they are known for the signal
and not for the background. When the corrections are known, they are often not incorporated
in the event generator tools that are employed. 
Except where noted, we have adopted a conservative approach and use calculations that
are only to lowest order. Almost all higher order corrections increase the rates, and are sometimes
included by multiplying the lowest order rates by a so-called K-factor.
These corrections are typically $\sim 1.5$ and can occasionally be as large as 2.0.
Since the corrections depend on kinematical
details this procedure is at best an approximation. Uncertainties from the choice of scale and
structure functions are at the 20\% level except in cases involving the production of very light
states. The total cross-section for b-quark production is particularly uncertain.

The level of simulation used in the processes varies quite widely.
For a few processes a full GEANT\cite{geant} style simulation has 
been carried out.
Such simulations are very slow ($\sim few \times 10^5$ Mips/event)
and are difficult to carry out for processes where large number of
events need to be simulated and many strategies for extracting signals 
need to be pursued. In these cases a particle level simulation and
parameterized detector response is used. A lower level of simulation
involving partons ({\it i.e.} leptons and jets) and parameterized response
is fast  and might be required when the underlying parton 
process is not present in full event generators. This last level of
simulation is useful for exploring signals but often leads to
overly optimistic results, particularly where the reconstruction
of invariant masses of jets are involved. None of the results
included here use this last level of simulation, unless stated explicitly.   

\subsection{Detectors}

Two large, general-purpose $pp$ collider detectors will be constructed for
LHC:  ATLAS\cite{atlas} and CMS\cite{cms}.  
Both collaborations completed Technical Proposals for
their detectors in December 1994, and were formally approved
in January 1996.  Though they differ in most details, the detectors 
share many common emphases which derive from the physics goals of LHC:
\begin{itemize}
\item they both include precision electromagnetic calorimetry;
\item they both use a rather ambitious magnet (though of different geometries)
in order to obtain good muon identification and precision momentum 
measurement;
\item both have lepton identification and measurement over $|\eta| < 3$;
\item
they both incorporate ambitious multi-layer silicon
tracker systems for heavy flavor tagging (the usefulness of this capability is
an important lesson from the Tevatron); 
\item
they both include forward calorimetry for large $\eta$ coverage in order to
obtain the required $\met$ resolution.
\end{itemize}

\begin{figure}
\epsfxsize=8.5cm
\dofig{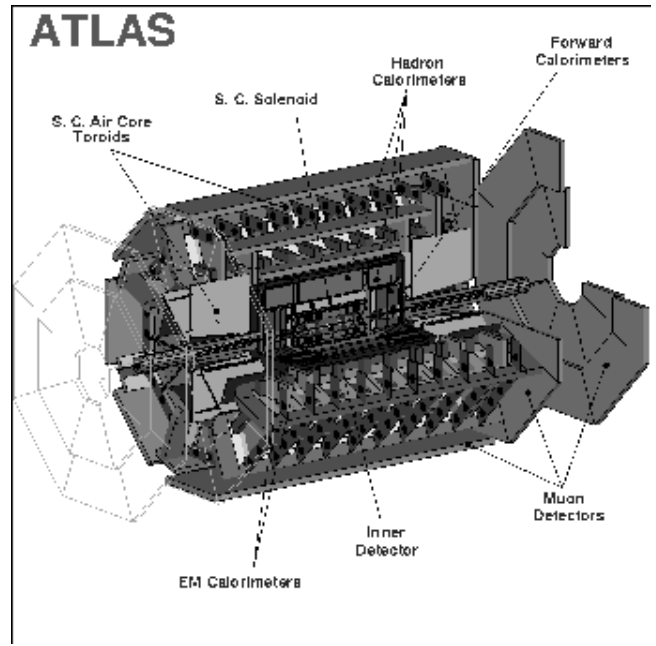}
\caption[]{The ATLAS detector
\label{atlasdetector}}
\end{figure}

The ATLAS detector is shown in Fig.\ref{atlasdetector}.
It uses a tracking system employing silicon pixels, silicon strip
detectors, and a transition radiation tracker, all contained within a
superconducting solenoid.  The charged track resolution is $\Delta p_T/p_T=20\%
$ at $p_T=500\,$GeV/c.  The tracker is surrounded by an electromagnetic
calorimeter using a lead-liquid argon accordion design; the EM calorimeter
covers $|\eta|<3$ (with trigger coverage of $|\eta|<2.5$)
and has a resolution of $\Delta E/E = 10\%/\sqrt{E} \oplus 0.7\%$.  The hadronic
calorimeter uses scintillator tiles in the barrel, and liquid argon in the
endcaps ($|\eta|>1.5$); its resolution is $\Delta E/E = 50\%\sqrt{E} \oplus 3\%$.  
Forward calorimeters cover the region $3 < |\eta| < 5$ with a resolution 
$\Delta E/E = 100\%\sqrt{E} \oplus 10\%$.  Surrounding the calorimeters 
is the muon system.  Muon trajectories are measured using three layers
of chambers (MDT's and CSC's) in a spectrometer using a large air-core toroid 
magnet.  The resulting muon momentum measurement is $\Delta p_T/p_T=8\%$ at
$p_T=1\,$TeV/c and  $\Delta p_T/p_T=2\%$ at $p_T=100\,$GeV/c.  Muons may be
triggered on over the range $|\eta|<2.2$.

\begin{figure}
\epsfxsize=8.5cm
\dofig{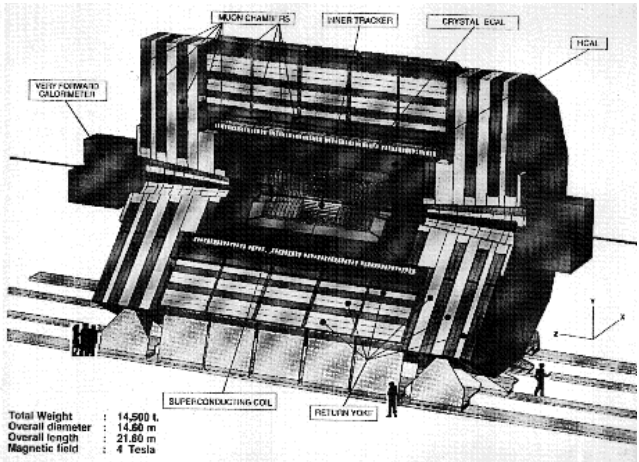}
\caption[]{The CMS detector
\label{cmsdetector}}
\end{figure}

The CMS detector is shown in Fig.\ref{cmsdetector}.
The tracking system is based on silicon pixels, silicon strip
detectors, and microstrip gas chambers.  
The charged track resolution is $\Delta p_T/p_T=5\%$ at $p_T=1\,$TeV/c
and $\Delta p_T/p_T=1\%$ at $p_T=100\,$GeV/c.
CMS has chosen a precision electromagnetic calorimeter 
using lead tungstate (PbW0$_4$) crystals, covering 
$|\eta|<3$ (with trigger coverage of $|\eta|<2.6$).
Its resolution at low luminosity
is $\Delta E/E = 2\%\sqrt{E} \oplus 0.5\%$.  The surrounding
hadronic calorimeter uses scintillator tiles in the barrel and
endcaps; its resolution is $\Delta E/E = 65\%\sqrt{E} \oplus 5\%$.  The
region $3 < |\eta| < 5$ is covered by 
forward calorimeters using parallel-plate chambers or quartz fibers 
and having a resolution of about 
$\Delta E/E = 130\%\sqrt{E} \oplus 10\%$.  The calorimeters 
are contained in a 4~tesla superconducting coil which provides the magnetic field
for charged particle tracking.  
Muon trajectories outside the coil are measured in four layers
of chambers (drift tubes and CSC's) embedded in the iron return yoke.  
The muon momentum measurement using the muon chambers and the
central tracker covers the range 
$|\eta|<2.4$ with a resolution $\Delta p_T/p_T=5\%$ at
$p_T=1\,$TeV/c and  $\Delta p_T/p_T=1\%$ at $p_T=100\,$GeV/c.  The muon
trigger extends over $|\eta|<2.1$.

Significant contributions to both detectors are planned to be made by U.S.
groups.  For ATLAS, these groups involve about 200 physicists and engineers
from 27 U.S. institutions; for CMS, about 300 physicists and engineers from
37 U.S. institutions.  Contributions to ATLAS include one half to one third of
the silicon pixels, one third to one quarter of the silicon strips, and the
barrel transition radiation tracker; all or part of the readout for the liquid argon
calorimeter, the EM section of the forward calorimeters, and about one third of
the scintillator tile calorimeter; the endcap muon system, and contributions to
the level 1 and level 2 triggers.   For CMS, the list includes the forward
silicon pixels, the hadron calorimeter system (management of the whole
project and construction of the barrel and forward calorimeters);
the EM calorimeter front-end; the endcap muon detectors
(management of the system)
and contributions to the level 1 and level 2 triggers
(including management of the calorimeter trigger).
At the time of writing (June 1996) negotiations are still ongoing between CERN
and the U.S. funding agencies over the level of financial contribution to be
made to ATLAS and CMS.  Until final figures are arrived at, the contributions
of U.S. groups are of course subject to revision.  

One important, but less tangible, contribution from the U.S. groups is 
their involvement in the Tevatron collider program with the CDF and D\O\
experiments.  
These provide a unique opportunity to learn, in a somewhat less
demanding environment, how to deal with many of the
challenges of high luminosity
hadron collider physics, such as energy from pileup events,
discrimination between multiple vertices, trigger rates dominated by 
backgrounds, and heavy flavor tagging, in a real experiment.

\section{Higgs Physics}

We will use ``Higgs bosons'' to refer to any scalar particles whose existence is connected
to electroweak symmetry breaking. Generically, Higgs bosons couple most strongly to heavy particles.
Their production cross section in hadron colliders is small resulting in final states with low
signal to background ratios. The ability to detect them and measure their mass
provides a set of benchmarks by which detectors can be judged. 
A specific model is required in order to address the quantitative questions of how well the
detector can perform. While one may not believe in the details of any 
particular model, a 
survey of them will enable general statements 
to be made about the potential of the LHC and its detectors.

\subsection{Standard Model Higgs}
\begin{figure}[t]
\vbox to 12.8cm{
\epsfxsize=8.5cm
\dofig{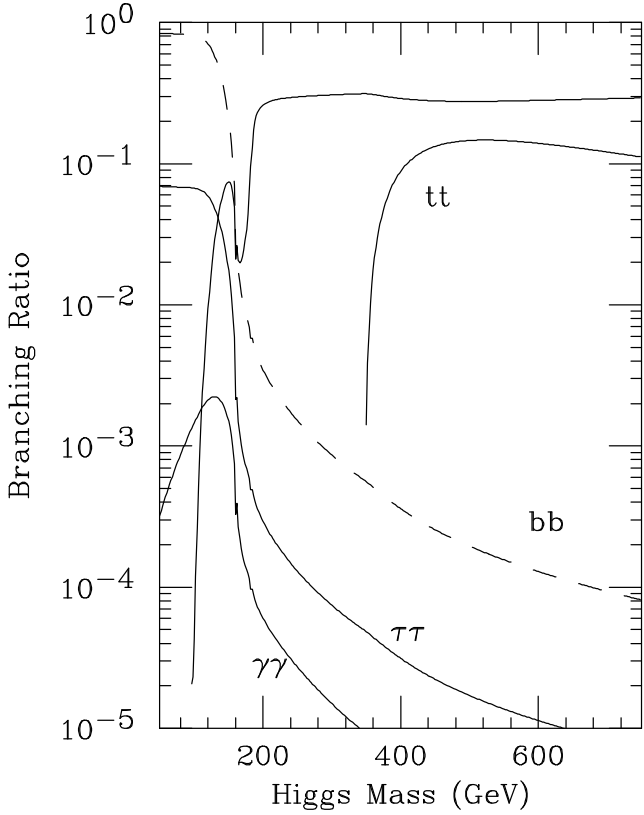}
\caption[]{The branching ratios of the 
standard model Higgs boson as a function of its mass.
The highest lying curve at large mass is the $ZZ$ final state. Not shown is the $WW$
rate which makes up almost all of the unaccounted for branching ratios.
\label{higgsbr}}
}
\end{figure}

All the properties of the standard model Higgs boson are determined once its mass is fixed.
The search strategy at LHC is therefore well defined.
The current limit on the mass of the Higgs boson is 
$M_H>65$ GeV for experiments at LEP\cite{higgslimit}.
Before the LHC gives data, masses up to 95~GeV will have been excluded 
or discovered by LEP\cite{lepstudy}.
There are several relevant production mechanisms; $gg\to H$ via an intermediate quark or gauge boson loop;
$q\overline{q}\to WH$; $gg\to t\overline{t}H$; $gg\to b\overline{b}H$ and  $qq\to qqH$.
The relative importance of these processes depends upon the Higgs mass, the first dominates at small mass and the last at high
masses. The branching ratios are shown in Fig.~\ref{higgsbr}.

\subsubsection{$H \to \gamma \gamma$ and associated production channels} 

At masses just above the range probed by LEP, the dominant decay of the Higgs boson is to the
$b\overline{b}$ final state which is difficult to reconstruct. The decay to $\gamma\gamma$ is the most promising 
in this region. The branching ratio is very small and there is a large background from the pair production  of photons via
$q\overline{q}\to \gamma\gamma$, 
$gg \to \gamma\gamma$, and the bremsstrahlung process
$qg \to q(\to\gamma)\gamma$. Excellent photon energy resolution is required to 
observe this signal, and this process is one that drives the very 
high quality electromagnetic calorimetry of both experiments. 

CMS has a mass resolution of order 540 (870) MeV  at $m_H=110$ for 
low (high) luminosity\cite{cmsgamgam}.
The  mass resolution is worse at high luminosity due to event pile up and the presence of a preshower detector that is used to 
determine the photon direction. This preshower is necessary as there are multiple interactions and the primary vertex is 
not readily recognised. The preshower enables the photon direction to be determined with a precision of $40mr/\sqrt{E}$
and used to resolve the ambiguity in which of the several events contains the signal and
therefore what point along the beam is used in computing the diphoton invariant mass. It is not present at low luminosity.
The ATLAS mass resolution at high (low) luminosity is 1.2 (1.1) GeV for at $M_H=110$ GeV.
However the photon acceptance and identification efficiency are 
higher in the ATLAS  analysis\cite{atlasgamgam}, 
partly because CMS rejects photons
that convert in the inner detector.

In addition to the background from $\gamma\gamma$ final states,
there are $jet-\gamma$ and $jet-jet$ final states, that are much larger. A $jet/\gamma$ rejection factor of $\sim 10^3$ is needed to
bring these backgrounds below the irreducible $\gamma\gamma$ background. 
A detailed GEANT based study of the ATLAS detector has been performed to study these 
backgrounds\cite{atlasgamgam}.
Jets were rejected by applying cuts on the hadronic leakage, photon isolation and the measured width of the electromagnetic shower.
These cuts result in an estimate of these backgrounds which is a factor of six
below the irreducible $\gamma\gamma$ background. 
The background $Z\to e^+e^-$ where
both electrons are misidentified as photons was found to be significantly below 
the jet background, except in the range $m_H \sim m_Z$.  In this case, stringent track
rejection is needed. Given the uncertainties in the 
rates for these ``reducible'' backgrounds one can be confident that they are smaller than the irreducible $\gamma\gamma$
background, but they may not be negligible.

The CMS analysis for this process is as follows\cite{cms,cmsgamgam}. 
Two isolated photons are required one of which has $p_T>25$ GeV and the other has
$p_T>40$ GeV. Both are required to satisfy $\abseta <2.5$. Isolation means that there is no track or additional electromagnetic
energy cluster with $p_T>2.5 GeV$ in a cone of size $\Delta R=0.3$ around the photon direction.
The Higgs signal then appears as a peak over the smooth background. 
The signal to background ratio is small, but there are many
events. A curve can be fitted to the smooth background and subtracted from
data. 
Fig.~4 %\ref{fighiggam} 
shows the result of this
subtraction. Peaks are shown corresponding to Higgs masses of 90, 110 and 130
GeV. The figure also shows the
event rate needed to establish a signal of some significance as a function of the mass. From this one can see that
this mode can discover the Higgs if its mass is too high to be detected at LEP and below 
about 140 GeV. At larger masses the
branching ratio becomes too small for a signal to be extracted. The large event rate for this process implies that it becomes
effective for a more limited range of Higgs masses once the integrated luminosity 
exceeds $\sim 10$~fb$^{-1}$.
Results of the ATLAS study are similar and the reach of the two experiments is
similar\cite{atlasgamgam}

\begin{figure}
\epsfxsize=8.5cm
\dofig{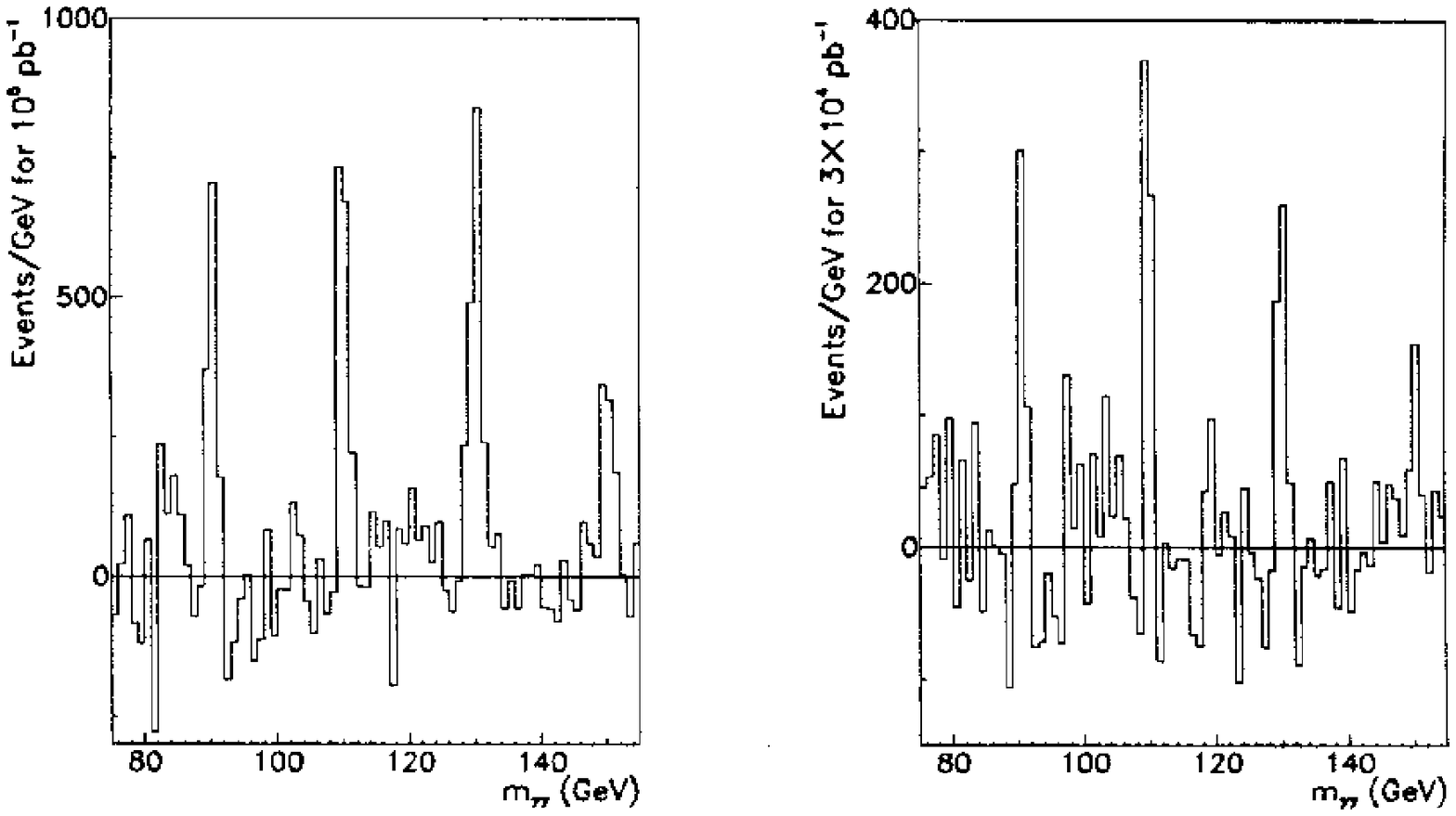}
\epsfxsize=8.5cm
\dofig{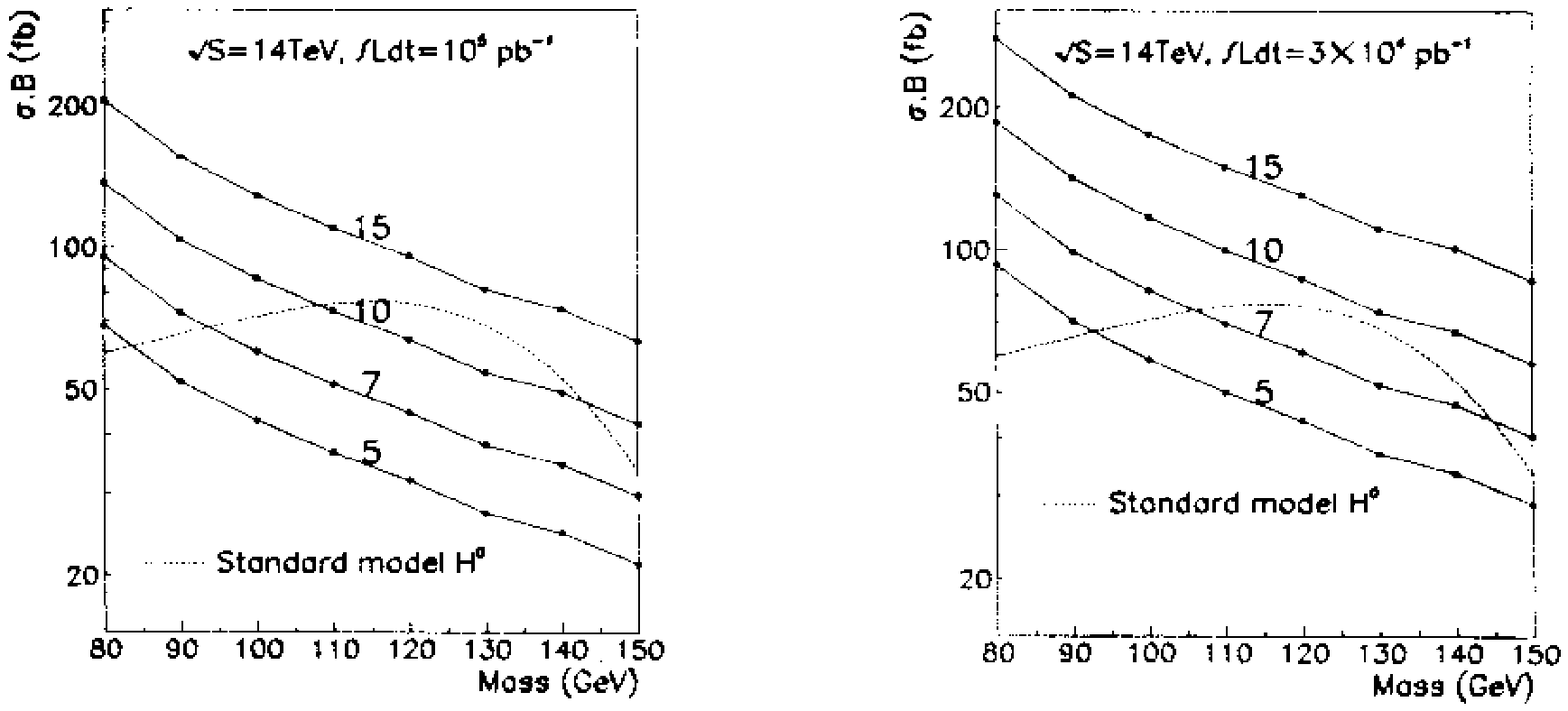}
\caption[]{(a)The invariant mass distribution of $\gamma\gamma$ pairs as simulated by the CMS collaboration. A smooth background has
been fitted and subtracted. (b) Curves showing the statistical significance of the peak as a function of the
Higgs mass and the product of production cross-section and branching ratio $\sigma B$. The dotted line corresponds to the
value of $\sigma B$ as a function of the standard model Higgs boson mass. The left (right) figures correspond to low (high)
luminosity running.
\label{fighiggam}}
\end{figure}
 
Another process is available at the lower end of the mass range. If the Higgs is produced in association with at $W$ or
$t\overline{t}$, the cross section is substantially reduced, but the presence of additional particles proportionally larger
reduction in the background.  Events are required to have an isolated lepton arising from the decay of the $W$ (or top quark).
This lepton  can be used to determine the vertex
position.
The process is only useful at high luminosity, for $10^5$ pb$^{-1}$
there are approximately 15 signal events for Higgs masses between 90 and 120 GeV (the falling cross-section is compensated by the
increased branching ratio for $H\to \gamma\gamma$)
 over an approximately equal background \cite{atlaswh,atlas}. The process will therefore provide confirmation of a discovery
made in the $\gamma\gamma$ final state without an associated lepton.

\subsubsection{$H \to b \overline b$}
The dominant decay of a Higgs boson if its mass is below $2M_W$ is to $b\overline{b}$. The signal for a Higgs boson produced in
isolation is impossible to extract. There is, as yet, no conceivable trigger for the process and the background production of
$b\overline{b}$ pairs is enormous. The production of a Higgs boson in association with a W or $t\overline{t}$ pair can provide a
high $p_T$ lepton that can be used as a trigger. A study was conducted by ATLAS of this very challenging channel\cite{froid95}.
Events were triggered by requiring a muon (electron) with $\abseta <2.5$ and $p_T>6 (20)$  GeV. A study was carried out of the
tagging efficiency to be expected for jets containing b-quarks\cite{atlastag}. $t\overline{t}$ events were generated and used both as a source of
b-jets and light quark jets. At low luminosity, the ATLAS detector has a layer  (the
so-called B-layer) at $\sim 5$~cm from the beam. 
In this case a b-tagging efficiency of 70\% is achieved 
with a rejection factor of 100 against light
quark jets.  The situation is somewhat worse at high luminosity as is  shown in
Fig.~5
%\ref{atlasbtag}.
This b-tagging efficiency is not significantly larger than that obtained by CDF\cite{cdftag}.

\begin{figure}
\epsfxsize=8.5cm
\dofig{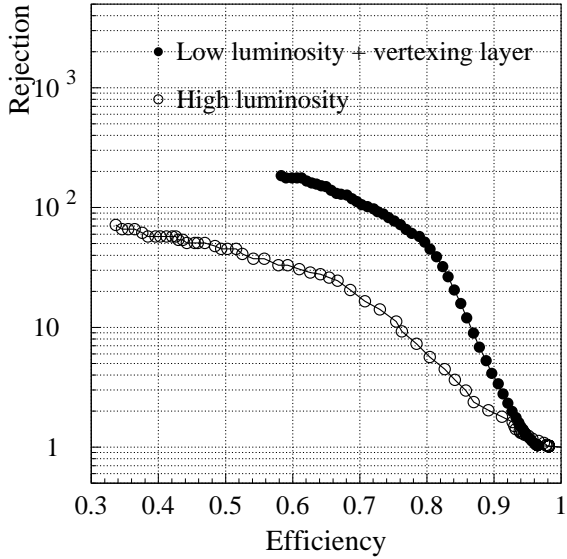}
\caption[]{Rejection factor for light quark jets as a function of the tagging efficiency for b-quark jets in the ATLAS
detector\cite{atlas}.
\label{atlasbtag}}
\end{figure}

The study of $H\to b\overline{b}$ assumed an
efficiency of 50\% and a background rejection of 100. Using this 
assumption the background from $Wb\overline{b}$ events is slightly
larger than that from $W$'s in association with light quark jets. 
The Higgs search is then limited by the background from real
b-quarks which is detector independent. Jets were retained if they 
had $p_T>16$ GeV and $\abseta <2.5$. In order to reduce  the
background from $t\overline{t}$ events a veto was applied to reject events 
with a second isolated lepton $p_T>6$ GeV and $\abseta
<2.5$ and additional jets with $p_T>15$ GeV in $\abseta <5$. For a 
luminosity of $10^4$ pb$^{-1}$, there are 175, 110 and 47 signal
events for Higgs masses of 80, 100 and 120 GeV from the $WH$ process. 
The reconstructed $b\overline{b}$ mass distribution is not gaussian, 
it has a tail on the low side.
Nevertheless fit to a gaussian  gives $\sigma\sim 11 $ GeV for a 
mass of 100 GeV. The position of the peak is also shifted down by
about 20\%. These two degradations are caused mainly by 
gluon radiation off the final state $b$ quarks and losses due to decays. 
The background arising from $Wb\overline{b}$ events is large;
approximately 3000, 2500 and 1880 events in a 
bin of width 40 GeV centered on the reconstructed $b\overline{b}$ mass peak.   
An additional background from $WZ(\to b\overline{b})$ is present if $m_H\sim M_Z$ and contributes an event rate approximately equal
to that of the signal.  The final state $t\overline{t}H(\to b\overline{b})$ was also studied. A third tagged b-jet was required. The signal and background
rates were similar to the $WH$ case\cite{froid95}.
 From this study we can draw the following conclusion. Extraction of a signal will be possible if at all
only over a very  limited mass range ($\sim 80-110$~GeV) and depends 
critically upon the b-tagging efficiency and background rejection.
The signal may be sufficient to confirm the discovery of a Higgs in another channel.

\subsubsection{$H \to ZZ^* \to 4\ell$}

The search for the Standard Model Higgs relies on the four-lepton channel over
a broad mass range from $m_H \sim 130\,$GeV to $m_H \sim 800\,$GeV.
Below $2m_Z$, the event rate is small and the background reduction more
difficult, as one or both of the $Z$-bosons are off-shell.  In this mass region
the Higgs width is small ($\simle 1\,$GeV) and so lepton energy or momentum
resolution is of great importance in determining the significance of a 
signal\cite{atlaszzstar}.

For $m_H < 2m_Z$, the main backgrounds arise from $t\overline t$, $Zb\overline
b$ and continuum $Z(Z/\gamma)^*$ production.  Of these, the $t\overline t$ background
can be reduced by lepton isolation and by lepton pair invariant mass cuts.
The $Zb\overline b$ background cannot be reduced by a lepton pair invariant
mass cut but can be suppressed by isolation requirements.  The $ZZ^*$
process is an irreducible background.
Both CMS and ATLAS studied the process for $m_H = 130$, 150 and 170~GeV.  
Signal events
were obtained from both $gg \to H$ and $WW/ZZ$ fusion processes, giving
consistent cross sections 
$\sigma\cdot B \approx 3$, 5.5 and 1.4~fb respectively (no $K$-factors being
included). 

In the CMS study\cite{cms,cmszzstar}
event pileup appropriate to 
${\cal L}=10^{34}\,{\rm cm}^{-2}{\rm s}^{-1}$ was modelled by superimposing 15
minimum bias events (simulated by QCD dijets with $p_T \geq 5\,$GeV/c).
The muon resolution was obtained from a full simulation of the detector
response and track-fitting procedure.  This was then parameterized as a function
of $p_T$ and $\eta$.  Internal bremsstrahlung was generated using the PHOTOS
program and leads to about 8\% of reconstructed $Z \to \mu^+ \mu^-$ pairs 
falling outside a $m_Z \pm 2 \sigma_Z$ window for $m_H = 150\,$GeV.  
The reconstructed $\mu^+ \mu^-$ mass has a resolution $\sigma_Z = 1.8\,$GeV in
the Gaussian part of the peak.  
The electron resolution was obtained from a detailed GEANT simulation of the
calorimeter, including the effects of material in the beampipe and the tracker,
and the reconstruction of electron energy in the crystal calorimeter.
Including internal and external bremsstrahlung, and using a $5\times 7$ crystal
matrix to reconstruct the electron, the mass resolution $\sigma_Z = 2.3\,$GeV
and the reconstruction efficiency is about 70\% (within $m_Z \pm 2 \sigma_Z$).

Events were selected which had one electron with $p_T > 20\,$GeV/c, one with
$p_T > 15\,$GeV/c and the remaining two with $p_T > 10\,$GeV/c, all within
$|\eta| < 2.5$.  For muons, the momenta were required to exceed 20, 10 and
5~GeV/c within $|\eta| < 2.4$.  One of the $e^+e^-$ or $\mu^+\mu^-$ pairs was
required to be within $\pm 2\sigma_Z$ of the $Z$ mass.  This cut loses
that fraction of the signal where both $Z$'s are off-shell, about a 24\%
inefficiency at $m_H = 130\,$GeV and 12\% at $m_H = 170\,$GeV.
The two softer leptons were also required to satisfy $m_{\ell\ell} > 12\,$GeV.
Additional rejection is obtained by requiring 
that any three of the four leptons be isolated in the tracker, demanding that
there is no track with $p_T > 2.5\,$GeV/c within the cone $R<0.2$ around the
lepton.  This requirement is not very sensitive to pileup as the $2.5\,$GeV/c
threshold is quite high.  This yields signals at the 
level of 7.4, 15.2 and 5.0 standard deviations for
$m_H = 130$, 150, and 170~GeV in $2\times 10^5\,{\rm pb}^{-1}$.  

The ATLAS\cite{atlas,atlaszzstar} study followed a similar technique.  
The detector resolutions and reconstruction efficiencies were obtained using
detailed detector simulations, including the effects of pileup.  
For the four-electron mode, the Higgs mass resolution at $m_H = 130\,$GeV
is 1.7~(1.5)~GeV at high (low) luminosity, including the effect of electronic 
noise in the calorimeter.  For muons, the
corresponding figure is 2.0~GeV after correcting for muon energy losses in the
calorimeter; this can be improved to about 1.6~GeV by combining the muon
momentum measured in the muon system with that obtained from the central
tracker after the tracks have been matched.  
Events were selected which had two leptons with $p_T > 20\,$GeV/c, and 
the remaining two with $p_T > 7\,$GeV/c, all within
$|\eta| < 2.5$.  One of the $e^+e^-$ or $\mu^+\mu^-$ pairs was
required to be within $\pm 6\,$GeV of the $Z$ mass.  
The two softer leptons were also required to satisfy $m_{\ell\ell} > 20\,$GeV.

ATLAS used a combination of calorimeter isolation and impact parameter
cuts.  The isolation criterion is that the transverse energy
within $R=0.3$ of the lepton be less than $E_T^{cut}$.  Values of $E_T^{cut}$
of 3, 5, and 7~GeV were used for $4\mu$, $ee\mu\mu$ and $4e$ modes at
$10^{33}$
($10^{34}$) luminosity to obtain a constant signal efficiency of 85\% (50\%).
Tighter cuts can be used for muons because they do not suffer from
transverse leakage of the EM shower.  
The impact parameter, as measured in
the silicon tracker, is used to further reduce the background from heavy flavor
processes ($t \overline t$ and $Z b \overline b$)\cite{atlaszzstar}.  
ATLAS obtain signals at the 
level of 8.5 (7.8), 22 (18) and 6.5 (5) standard deviations for
$m_H = 130$, 150, and 170~GeV in 
$10^5\,{\rm pb}^{-1}$ $(3\times 10^4\,{\rm pb}^{-1})$.  
The four-lepton mass distributions are shown in Fig.~6. %\ref{figzzstar}.

\begin{figure}
\vbox to 11.5cm{
\epsfxsize=8.5cm
\dofig{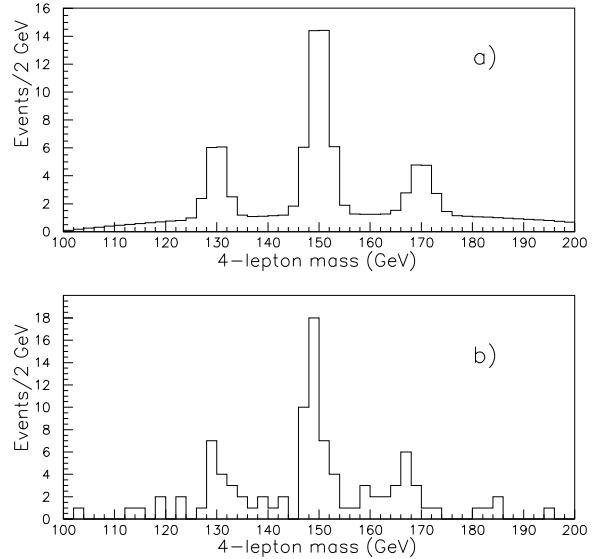}
\caption[]{Reconstructed four-lepton mass above background, 
for $m_H = 130$, 150 and 170~GeV, and an integrated luminosity of
$3\times 10^4\,{\rm pb}^{-1}$ (low luminosity) as simulated by the ATLAS collaboration.  $(a)$ indicates the expected average 
number of events; $(b)$ shows the result of one experiment, obtained with
randomized statistics in each mass bin.
\label{figzzstar}}
}
\end{figure}

\subsubsection{$H \to ZZ \to 4\ell$}

The $H \to ZZ \to 4\ell$ channel is sensitive over a wide range of Higgs
masses from $2m_Z$ upwards: to about 400~GeV with $10^4\,{\rm pb}^{-1}$ and
to about 600~GeV with $10^5\,{\rm pb}^{-1}$.  For lower Higgs masses, the width
is quite small and precision lepton energy and momentum measurements are
helpful; for larger masses the natural Higgs width becomes large.
The main background is continuum $ZZ$ production.

CMS\cite{cms,cmszzstar} studied the process for $m_H = 300$, 500 and 600~GeV.  
The electron and muon resolutions and the selection cuts were the same as used
for the $ZZ^*$ channel.  Internal and external bremsstrahlung were simulated
using the PHOTOS program and a GEANT detector simulation. Two  $e^+e^-$ or
$\mu^+\mu^-$ pairs with a mass within $\pm 6\,$GeV of $m_Z$ were required.
No isolation cut was imposed as the remaining backgrounds are small.
The resulting 4-lepton invariant mass distributions are shown in Fig.~7.
%~\ref{figzz}.  
With $10^5\,{\rm pb}^{-1}$ a signal in excess of six standard
deviations is visible over the entire range $200 < m_H < 600\,$GeV.  
ATLAS obtains very similar results\cite{atlas}.  

\begin{figure}[t]
\vbox to 13cm{
\epsfxsize=8.5cm
\dofig{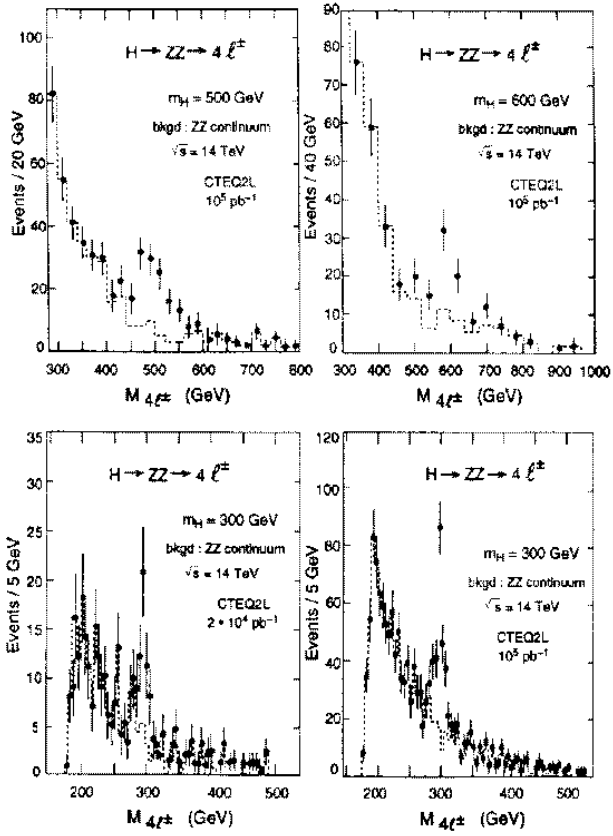}
\caption[]{Mass distribution in $H\to ZZ^*\to 4\mu$ for $M_H=150$ GeV as simulated by CMS including all bremsstrahlung losses.
\label{figzz}}
}
\end{figure}

\subsubsection{$H \sim 1$~TeV ($\ell\ell\nu\nu$, $\ell\ell$jj, $\ell\nu$jj, etc.)}

As the Higgs mass is increased further, its width increases and the production rate falls and one must turn to decay
channels that have a larger branching ratio. The first of these is $H\to ZZ\to\ell\ell\nu\overline{\nu}$.
Here the signal involves looking for a $Z$ decaying to lepton pairs and a large amount of missing energy. The signal appears
as a Jacobian peak in the missing $E_T$ spectrum.  There are more potentially important sources of background in this channel than
in the $4\ell$ final state. In addition to the irreducible background from $ZZ$ final states, one has to worry about $Z+jets$
events
where the missing $E_T$ arises from neutrinos in the jets or from cracks and other detector effects that cause jet energies to be
mismeasured. At high luminosity the background from the pile up of minimum bias events produces a $\ETmiss$ spectrum that falls very
rapidly and is completely negligible for $\ETmiss>100$ GeV,
provided the calorimeter extends to $\abseta < 5$. ATLAS conducted 
\cite{bosman} a full GEANT  based study of this
background for which 5000 high transverse momentum $Z+jet$ events were fully simulated.
The events were selected so that a large fraction of them had jets going into the region $0.9<\abseta <1.3$ where
ATLAS has weaker jet energy resolution due to the crack between the endcap and barrel hadron calorimeters.
 The dominant part of the $Z+jets$ background that remains is that where the missing $E_T$ arises from the semi-leptonic
decays of b-quarks in the jets. The contribution from detector effects is not dominant.

\begin{figure}
\epsfxsize=8.5cm
\dofig{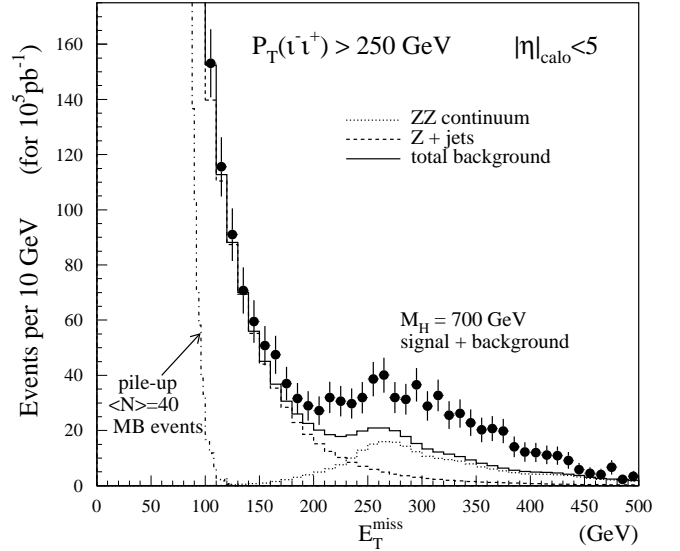}
\caption[]{Missing $E_T$ spectrum for the $H\to ZZ\to \ell\ell\nu\overline{\nu}$ process. The background contributions are shown
separately; $Z+jets$ (dashed); ZZ (dotted) and minimum 
bias pile up (dot-dashed). The signal due to a Higgs boson of mass 700 GeV.
\label{higgsllnunu}}
\end{figure}

Figure \ref{higgsllnunu} shows the missing $E_T$ spectrum at high luminosity ($10^5$ pb$^{-1}$).
On this plot the  $Z+jets$ background is estimated from a parton level
simulation as there are insufficient statistics in the full study to obtain the full missing $E_T$ spectrum. This estimate
correctly models the contribution from b-decays which the full study showed to be dominant. 
A cut was imposed requiring that
reconstructed $Z\to \ell\ell$ has $p_T(Z)>250$ GeV. This cut causes the $ZZ$ background to 
peak. This effect is less pronounced 
if a cut is made on $\ETmiss$ and then the plot is remade with $p_T(Z)$ on the abscissa.
The statistical significance of the signal shown is large but it is difficult to assess at this stage the true significance when
 data are actually taken.   The dominant $ZZ$ background has QCD corrections of order 40\% \cite{baur}. Once data are available
this background will be measured. The CMS analysis of this 
process\cite{cms,cmsllnunu} uses a central jet veto requiring that there be
no jets with $E_T>150$  GeV within $\abseta<2.4$. By requiring a jet in the far forward region (see below), most of the remaining
$ZZ$ background can be rejected. A study by CMS requiring a jet with $E>1 TeV$ and $2.4<\abseta <4.7$, produces an improvement of
approximately a factor of three in the signal to background ratio at the cost of some signal. This mode is only effective for high mass higgs bosons and
becomes powerful only at high luminosity. Nevertheless it will
provide an unambiguous signal.

Substantially larger event samples are available if the decay modes $H\to WW\to \ell\nu +jets$ and $H\to ZZ\to \ell\ell +jets$
can be exploited efficiently. In order to do this one has to reduce the 
enormous $W+jets$ and $Z+jets$ background by kinematic cuts.
Henceforth the discussion will be for the $WW$ final state; the $ZZ$ state is similar.
The first step is to reconstruct the $W$ decay to jets\cite{zmushko}. A particle level simulation was used including the effects of
pile up at high luminosity. Basic calorimeter cells of 
$\Delta \phi \times \Delta\eta =0.05\times 0.05$ 
 and energy threshold of $E_T=1$ GeV per cell were used. 
Jets were found using a cone of size $\Delta R=0.5$  and required to have $E_T> 350$ GeV. Within these cones two smaller jets with
$\Delta R=0.2$ and $E_T>50$ GeV were reconstructed. This algorithm 
reconstructs $W\to jets$ with an efficiency of about 60\% and
a W mass resolution of approximately 6.5 GeV for $W's$ produced in the decay of 1 TeV Higgs bosons. 
The mass resolution is slightly better at low luminosity where pile up is unimportant. These cuts applied to
the $W(\to \ell\nu)+jets$ sample with $p_T(W)>200 $ GeV reduces the rate for this process by a factor of 500 and brings it to a
level approximately equal to that from $t\overline{t}$ production; $t\overline{t}\to Wb W\overline{b}$. A limited statics full
simulation of this method in the ATLAS detector is in qualitative agreement with the above study \cite{atlas}.

After these cuts, the backgrounds from $W+jets$ and $t\overline{t}$ are still larger than the signal from $H\to WW$ and topological
cuts are required.  One of the processes $qq\to Hqq$ produces the Higgs boson in association with jets at large rapidity.
These jets can be used as a tag to reject background. This forward jet tag will cause some loss of signal since the $gg\to H$
process lacks these forward jets. Hence it is only effective for high mass Higgs bosons where the $qq\to Hqq$ process is a significant part
of the cross section. The central part (in rapidity) of the Higgs events is expected to have less jet activity in it than the
background, particularly that background from $t\overline{t}$.  
At low luminosity, requiring that the events have no additional jets (apart from the
ones that make up the W candidate)  with $E_T>15$ GeV and
$\abseta< 2$ loses approximately 30\% of the signal and reduces the 
background from $W+jets$ ($t\overline{t}$) by a factor of 3
(30). At high luminosity the requirement has to 
be raised to $E_T>40$ GeV  in order to preserve the efficiency
for the
signal. The rejection factors for $W+jets$ and $t\overline{t}$ are then 2.5 and 12.

The forward jet tagging was investigated in ATLAS as follows. Clusters energy of size $\Delta R=0.5$ were found in the
region $2<\abseta <5$. Events from the pile up of minimum bias events have jets in this regions so the threshold on $E_T$ of
the jet must be set high enough so that these jets do not generate tags in the background. If the individual calorimeter cells are
required to have $E_T>3$ GeV, then there is there is a 
4.6\% ( 0.07\%) probability that the pile up at high luminosity  will
contribute a single (double) tag to an event that would otherwise not have one for tagging jets with $E_T>15$ GeV and $E>600$ GeV.
The requirements for single and double tags are then applied to the signal from a Higgs boson of mass 1 TeV and the various
backgrounds. The pile up contributions are included and the event rates for a
luminosity of $10^5$ pb$^{-1}$ shown in table \ref{taggingtable}. 

\begin{table}
\begin{center}
\begin{tabular}{|l|r|r|r|r|}
\hline
Process&Central&Jet&Single&Double\\
&cuts&veto&tag&tag\\
\hline\hline
$H\to WW$&364&251&179&57\\
\hline
$t\overline{t}$&5620&560&110&5\\
\hline
$W+jets$&9540&3820&580&12\\
\hline
pileup&&&160&2\\
\hline
\end{tabular}
\caption[]{$H \to WW \to \ell\nu jj$ signals and backgrounds,
for $m_H=1$~TeV, before and after cuts in the forward region (see text).
The rates are computed for an integrated luminosity of $10^{5}\,{\rm pb}^{-1}$
and a lepton efficiency of 90\%. Only the $qq \to Hqq$ contribution to the signal
is included. Table from an ATLAS simulation.
\label{taggingtable}}
\end{center}
\end{table}

It can be seen from the table that it may be 
possible to extract a signal but the quoted signal to background 
ratios should not be
taken too seriously. However, other kinematic quantities may be used to
further discriminate between the signal and the background.
The $ZZ$ final state is cleaner as there is no $t\overline{t}$ background 
but the event rates are much smaller. 

\begin{figure}[t]
\epsfxsize=8.5cm
\dofig{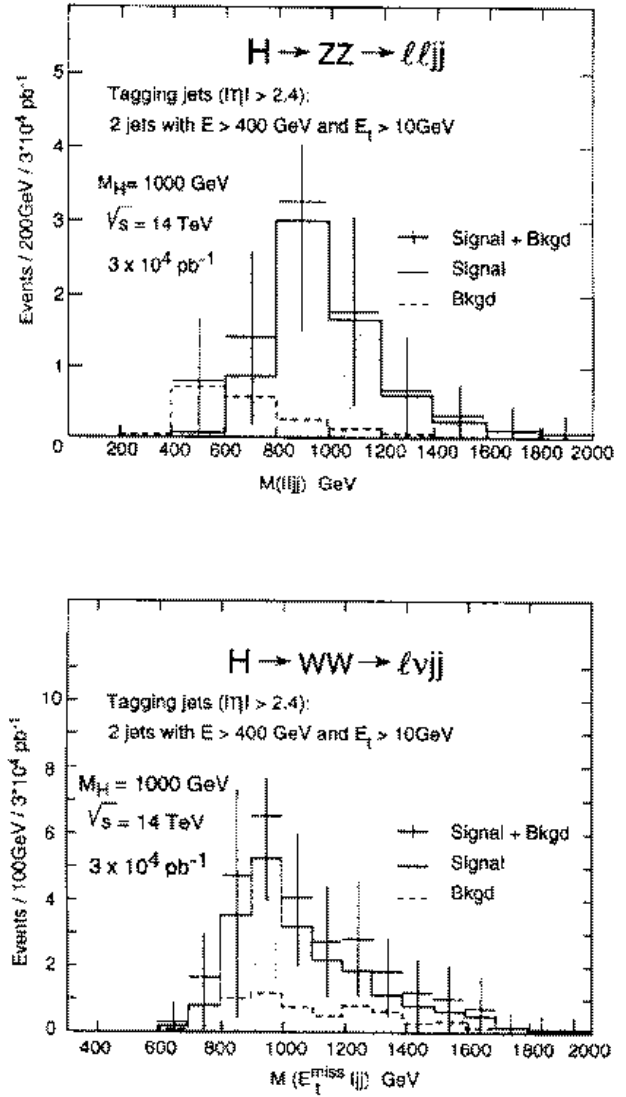}
\caption[]{Reconstructed diboson mass distributions in the final
states $\ell\ell jj$ and 
$\ell\nu jj$ showing a peak from a 1~TeV Higgs boson from a CMS simulation.
\label{cmszzjet}}
\end{figure}

A separate study was performed by the CMS group\cite{cms,cmslnujj}. Here two 
tagging jets with $\abseta >2.4$, $E_T>10$ GeV and $E>400$ GeV are
required. Two central jets are required with in 
invariant mass within 15 GeV of the $W$ or $Z$ mass. For the $ZZ$ case,
the $Z$ is reconstructed from $e$ or $\mu$ pairs with 
invariant mass within 10 GeV of the Z mass; each lepton has
$p_T>50$ GeV and the pair has $p_T>150$ GeV. For the $WW$ case, 
at least 150 GeV of missing $E_T$ is needed and the charged lepton
from the $W$ has $p_T>150$ GeV.  The result of this study is 
shown in Fig.~\ref{cmszzjet}. 

\subsubsection{Summary of standard model Higgs}
The LHC at full luminosity will be {\bf able to probe the entire range of allowed Higgs masses from the value reachable by
LEP up to the value where it is no longer sensible}  to speak of an elementary Higgs boson using final states that one is absolutely
confident will be effective: $\gamma\gamma$, $4\ell$ and $2\ell \nu\overline{\nu}$.
Additional final states that afford an excellent chance of having a signal will be exploited to support these; $b\overline{b}$ and
$\ell\nu+jets$, $\ell\ell +jets$. The failure to find a boson
over this range would therefore enable the standard model to be ruled out. 
The Higgs sector then either consists of non-standard Higgs
bosons or the electroweak symmetry breaking is via some strongly coupled process that will manifest itself in the study of $WW$
scattering. The next subsection is devoted to an example of the former type.

\subsection{SUSY Higgs}
As stated above the minimal supersymmetry model (MSSM) has three neutral and 
one charged Higgs bosons;  $h$, $H$, $A$ and $H^\pm$.  
These arise because supersymmetric models, unlike the standard model, need different
Higgs bosons to generate masses for the up and down type quarks. In the standard model one parameter,
the Higgs mass, is sufficient to fully fix its properties. In the Minimal supersymmetric model, two parameters are needed.
These can be taken to be the mass of $A$ which is unconstrained, and the ratio ($\tan \beta$) of the vacuum expectation values of
the higgs fields that couple to up-type and down-type quarks. If $\tan \beta$ is $O(1)$, then coupling of the top quark to Higgs
bosons ($\lambda_t$) is much larger than that of bottom quarks ($\lambda_b$) as is the case in the standard model. 

None of these Higgs bosons has been observed, so we need consider only the regions of parameter space not yet excluded.
The masses of $h$ and $H$ are given in terms of the mass of $A$ and $\tan \beta$. 
The charged Higgs boson  $H^\pm$ is heavier than $A$ ($M_{H^{\pm}}^2\sim M_A^2+M_W^2$). 
$H$ is heavier than $A$ and, at large values of $M_A$
$A$ and $H$, become almost degenerate. The mass of the lightest boson, $h$,
 increases with the mass of $A$ and reaches a plateau for $A$ heavier than
about 200GeV. The actual values depend on the masses of the other particles in the theory particularly the top quark
\cite{MSSMmasses}.
There is also a dependence (via radiative corrections) on the unknown masses of the other supersymmetric particles. This dependence
is small if these particles are heavy, so it is conventional to assume that this is the case. The only uncertainty in the masses of
the Higgs bosons then arises from the error on the top quark mass. 
Unfortunately the upper bound on the mass of $h$ is such that
it might be out of the range of LEP, which is 95~GeV for small $\tan\beta$ and 88~GeV
for large $\tan\beta$.

In the limit of large $A$ mass, the couplings of the Higgs bosons are easy to describe. The couplings of $h$ become like those of
the standard model Higgs boson. The raises the possibility that if $h$ is observed at LEP, it may not be possible to distinguish it
from those of the standard model Higgs boson. The couplings of $A$ and $H$ to charge 1/3 quarks and leptons are enhanced at large
$\tan \beta$ relative to those of a standard model Higgs boson of the same mass. However, $A$ does not couple to gauge boson pairs
and the coupling of $H$ to them 
is suppressed at large $\tan \beta$ and large $M_A$.  The decay modes used above in the case of the standard
model Higgs boson can also be exploited in the SUSY Higgs case. $h$ can be searched for in the final state $\gamma\gamma$, as the
branching ratio approaches that for the standard model Higgs in the large $M_A$ (decoupling) limit.

The decay $A\to \gamma\gamma$ can also be exploited. This has the advantage 
that, because $A\to ZZ$ and $A\to WW$ do not occur,
the branching ratio is large enough for the signal to be useable for values of
$M_A$ less than $2m_t$ \cite{snowgamgam}.
The decay $H\to ZZ^*$ can be exploited, but at large values of 
$M_H$ the decay $H\to ZZ$, which provides a very clear signal for the
standard model Higgs, is useless owing to its very small branching ratio,
The decays of $h\to b\overline{b}$ can also be exploited.

In addition to these decay channels, several other possibilities open up due to the larger number of Higgs bosons and 
possibly enhanced branching ratios. The most important of these 
are the decays of $H$ and $A$ to $\tau^+\tau^-$ and $\mu^+\mu^-$, $H \to hh$, $A \to Zh$ and
$A\to t\overline t$.

\subsubsection{$H/A \to \tau\tau$}

In the MSSM, the $H\to\tau^+\tau^-$ and $A\to\tau^+\tau^-$ rates are strongly
enhanced over the standard model if $\tan\beta$ is large, resulting in the possibility of 
observation over a large region of parameter space.  
The $\tau^+\tau^-$ signature can be searched for either in a lepton$+$hadron
final state, or an $e+\mu$ final state.  As there are always neutrinos to
contend with, mass reconstruction is difficult, and $E_T^{miss}$ resolution is
critical.  In ATLAS, at high luminosity this resolution is 
$\sigma(E^{miss}_{T,x})=\sigma(E^{miss}_{T,y})=1.1/\sqrt{\sum E_T}$.
Irreducible backgrounds arise from Drell-Yan tau pair
production, $t\overline{t}\to\tau\tau$ and $b\overline{b}\to\tau\tau$.
Both CMS\cite{cmstautau} and ATLAS\cite{atlastautau} have 
studied $\tau^+\tau^-$ final states using full
simulation.  

%
% CMS studies: go with what's in the TP for now even though there are
% new studies from Ritva K. in progress because no new plots yet
%
For the lepton$+$hadron final state, there are additional reducible 
backgrounds from events with one hard lepton plus a jet which is misidentified
as a tau.  In the CMS and ATLAS studies,
events were required to have one isolated lepton with $p_T >15-40$~GeV 
depending on $m_A$ (CMS) or $p_T >24 $~GeV (ATLAS)
within $|\eta|<2.0 (2.4)$ and one tau-jet candidate
within $|\eta|<2.0 (2.5)$. 
A lepton reconstruction efficiency of 90\% was
assumed by both ATLAS and CMS. 

CMS identified tau-jets by requiring $50 < E_T < 120$~GeV for
$m_A < 300$~GeV and $E_T> 60$~GeV for $m_A >300$~GeV.  Exactly one charged
track with
$p_T>25-40$~GeV/c was required within $R=0.1$ of the jet 
axis, and no tracks with
$p_T>2.5$~GeV/c %or calorimeter cell with $E_T>2.5$~GeV 
in the annulus between $R=0.1$ and $R=0.4$.  ATLAS required that the tau jet have
$E_T> 40$~GeV, that the radius of the jet computed from the EM cells only be
less than 0.07; that less than 10\% of its transverse energy be between
$R=0.1$ and $R=0.2$ of its axis; and again, that exactly one charged
track with $p_T>2$~GeV/c point to the cluster. 
The CMS and ATLAS  selections are
about 40\%(26\%) efficient for taus, 
while accepting only $1/100$ ($1/400$) of ordinary light quark and gluon jets.

\begin{figure}
\epsfxsize=8.5cm
\dofig{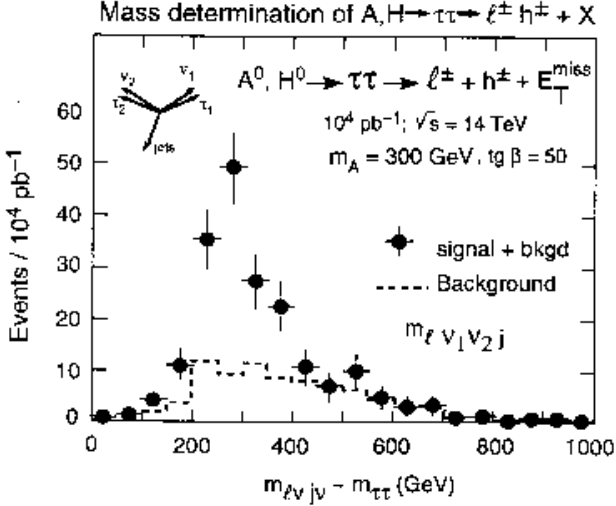}
\caption[]{ 
Invariant mass distribution of the $\ell j \nu\nu$ system for selected
events with $m_A=300$~GeV and $\tan\beta = 50$ from CMS.
\label{cmstautau}}
\end{figure}

CMS vetoed events having other jets with $E_T > 25$~GeV within $|\eta|<2.4$ (this
reduces the $t\overline{t}$ background); while ATLAS used cuts on
$E_T^{miss}$, the transverse mass of the lepton and $E_T^{miss}$, and
the azimuthal angle between the lepton and the tau-jet. 
The mass of the higgs may be
reconstructed by assuming the neutrino directions to be parallel to those of
the lepton and the tau-jet.  Resolutions of 12 and 14~GeV (Gaussian part) 
are obtained by ATLAS and CMS for $m_A=100$~GeV.  The reconstructed higgs peak
is shown in Fig.\ref{cmstautau}.

For the $e+\mu$ final state, CMS required a pair of opposite-sign 
unlike-flavor leptons with
$p_T > 20$~GeV/c and $|\eta|<2.0$.  There are large backgrounds
from the tau-pair processes listed above plus $WW$ production.  The $t\overline
t$ and $WW$ processes can be reduced to about one-fifth the Drell-Yan tau-pair
rate by a calorimeter circularity cut and by requiring $\Delta\phi$ between the
leptons be greater than 130$^\circ$.  The signal efficiency is about 40\%.
Both ATLAS and CMS find the sensitivity in the $e+\mu$ final state to be
less than for the lepton$+$hadron final state, owing to its smaller 
rate and less favorable decay kinematics.  

Taking the lepton$+$hadron and $e+\mu$ modes together, for the sum of
$h$, $H$ and $A$ decays, both ATLAS and CMS find that the large region 
of parameter space corresponding to
$\tan\beta \simge 6$ at $m_A = 125$~GeV rising to
$\tan\beta \simge 30$ at $m_A = 500$~GeV may be excluded at the 5$\sigma$
confidence level with $3\times 10^{4}\,{\rm pb}^{-1}$. 
ATLAS also finds some sensitivity to $\tan\beta \simle 2$ for
$125 < m_A <350$~GeV at very high integrated 
luminosities ($3\times 10^{5}\,{\rm pb}^{-1}$).  

\subsubsection{$H/A \to \mu\mu$}

The branching ratio for $H$ (or $A$) to $\mu^+\mu^-$ is smaller than that to $\tau^+ \tau^-$ by a factor of
$(m_{\mu} /m_{\tau} )^2$. The better resolution available in this
channel compensates to some extent for this  and the $\mu^+ \mu^-$ mode can be useful for large values of $\tan \beta$.
A signal of less statistical significance than that  in the $\tau^+ \tau^-$ could be used to confirm the discovery and make a 
more precise measurement of the mass and production cross section.
The ATLAS analysis \cite{froid96} requires two isolated muons with $p_T> 20 GeV$ and $\abseta <2.5$. The background from $t\overline{t}$ events
is rejected by requiring $\ETmiss <30 (60)$ GeV at low (high) luminosity. A jet veto could be employed to reduce this background
further, but this is ineffective at reducing the remaining dominant background for $\mu^+ \mu^-$ pairs from the Drell-Yan  process.
A cut on the transverse momentum of the muon pair, requiring it to be more than 10 GeV for small Higgs masses and 20 GeV for larger
masses reduces this background slightly. The remaining background is very large within $\pm 15 $ GeV of the Z mass. Above this
region the signal appears as a narrow peak in the $\mu^+\mu^-$ mass spectrum. In this troublesome region the signal will be
statistically significant if $\tan\beta$ is large enough but it appears as a shoulder on the edge of a steeply falling
distribution which may make it more difficult to extract a signal.

The significance of the signal in this channel is determined by the $\mu^+ \mu^-$ mass resolution and the intrinsic width of the 
Higgs resonance. The mass resolution in ATLAS is approximately $0.02m_A$ 
and $0.013m_A$ in CMS\cite{cmsmumu}.
At large $\tan \beta$, the masses of $A$ and $H$ are  almost degenerate and they cannot be resolved from each other.  The natural
width of $A$ is proportional to $\tan^2\beta$ and is approximately 3 GeV for $\tan\beta=30$ and $M_A=150$ GeV. The mode will
provide a $4\sigma$ signal for a region in the $M_A-\tan\beta$ plane covering $M_A>110$ GeV and $\tan\beta>15$ for an integrated
luminosity of $10^{5}$ pb$^{-1}$ .\footnote{The CMS event rates appear larger than the ATLAS
ones. CMS added the $A$ and $H$ rates whereas the ATLAS numbers correspond to the $A$ alone.}

\subsubsection{$A\to\gamma\gamma$}

The prospects for detecting the CP-odd Higgs boson ($A$) 
via its decay into photon pairs at the LHC were investigated at 
Snowmass\cite{snowgamgam}.
The CMS detector performance was adopted for a realistic study of 
observability.

Gluon fusion ($gg \to A$) via top and bottom quark 
triangle loop diagrams is the dominant production process if $\tan\beta \simle
4$; while for large $\tan\beta$ ($\simge 7$) $b$-quark fusion dominates. 
Both processes were included in this study. 
QCD corrections to the cross section were not included,
but the effect of QCD radiative corrections
on the branching fraction of $A \to b\bar{b}$ (which is reduced by about 
a factor of 2) was taken into account.  
For $\tan\beta \approx 1$ and 170 GeV $< m_A < 2m_t$ 
the branching fraction of $A \to \gamma\gamma$ is between
$5 \times 10^{-4}$ and $2 \times 10^{-3}$.

Events were simulated using 
PYTHIA 5.7 with the CTEQ2L parton distribution functions.
The backgrounds considered are QCD photon production, both the irreducible
two-photon backgrounds ($q \bar{q} \to\gamma \gamma $ and 
$ g g \to\gamma \gamma $) and 
the reducible backgrounds with one real photon
($q \bar{q} \to g \gamma$, $q g \to q \gamma$,  
and $g g \to g \gamma$). 
Both photons were required to have transverse energy ($E_T$) larger than 
40~GeV and $|\eta| < 2.5$.   
Both photons are required to be isolated, {\it i.e.}, 
(1) there is no charged particle in the cone in the cone $R = 0.3$; and 
(2) the total transverse energy $\sum E_T^{cell}$ 
is taken to be less than 5 GeV in the cone ring 0.1 $< R < 0.3$.
In this preliminary analysis, 
no rejection power was assumed against $\pi^0$'s with high $E_T$, and
all $\pi^0$'s surviving the cuts are considered accepted.
(This is very conservative and overestimates the background especially  
in the low mass $M_{\gamma \gamma}$ region.)

\begin{figure}[p]
\epsfxsize=8.cm
\dofig{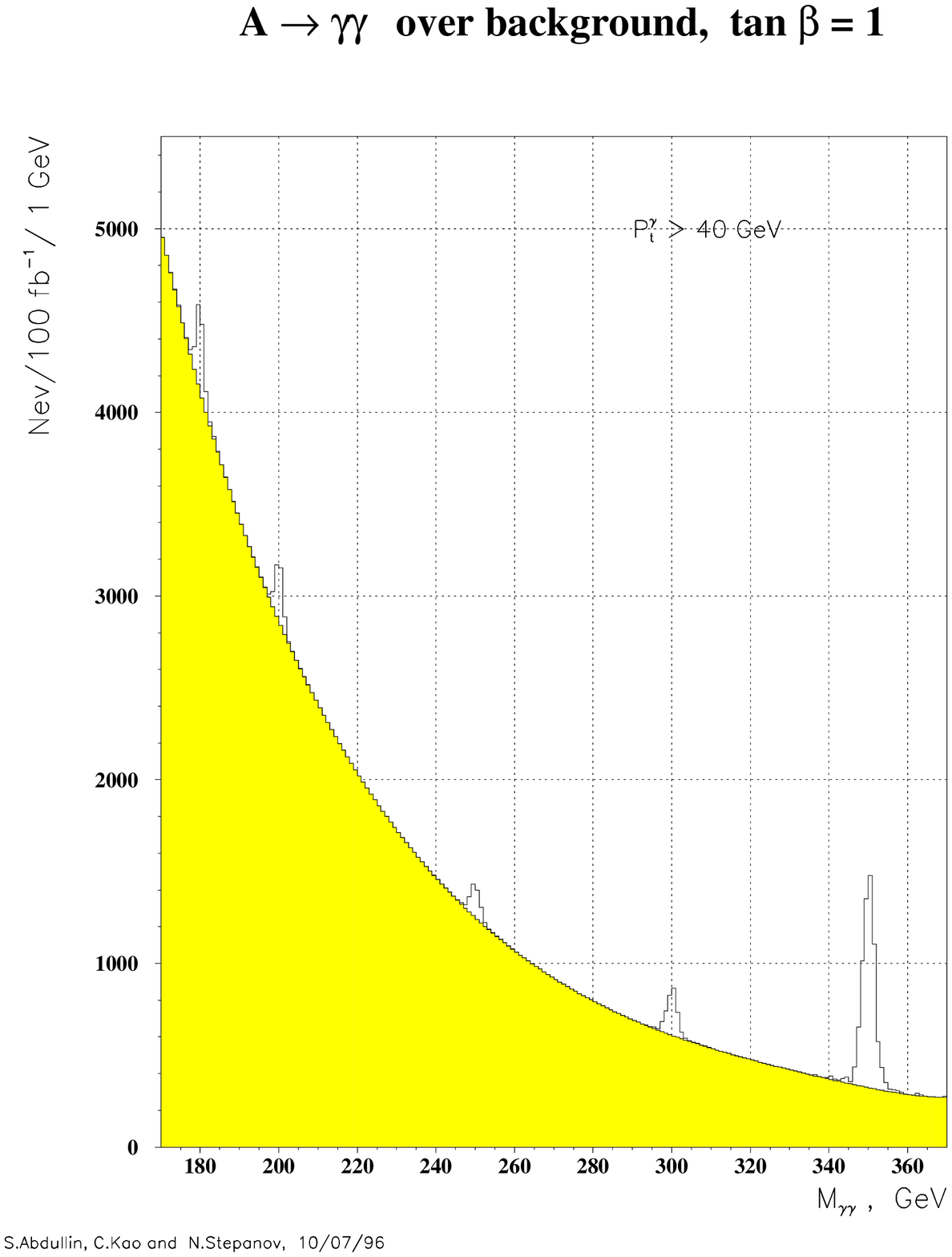}
\vspace{2cm}
\epsfxsize=8.cm
\dofig{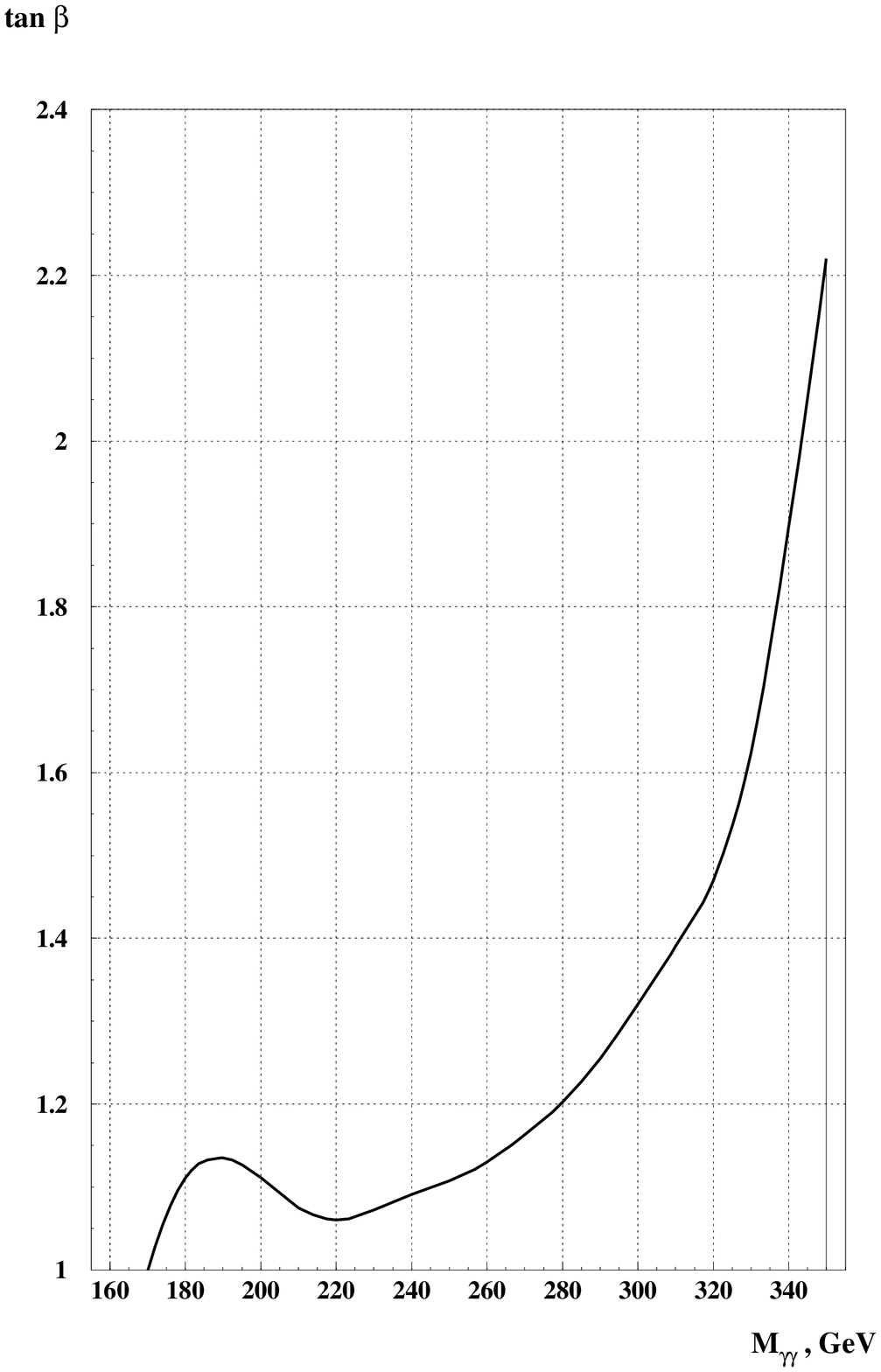}
%
%\vskip 12cm
%\special{psfile=a_to_gammagamma_fig1.ps
%          angle=0 hscale=40 vscale=40 hoffset=20 voffset=0}
%\special{psfile=a_to_gammagamma_fig2.ps
%          angle=0 hscale=40 vscale=40 hoffset=240 voffset=5}
%
\caption[]{$(a)$Number of $A\to\gamma\gamma$ events above background for
$100~fb^{-1}$ and $m_A = 180, 200, 250, 300$ and $350\,$GeV and $\tan\beta =1$.
$(b)$ $5\sigma$ discovery region for this channel in the $m_A-\tan\beta$ plane
for an integrated luminosity of $100~fb^{-1}$.
\label{atogamgam}}
\end{figure}

Fig.~\ref{atogamgam} shows the reconstructed $\gamma\gamma$ mass above
background for $m_A = 180, 200, 250, 300 and 350\,$GeV and $\tan\beta =1$.
The higgs peaks are clearly visible. 
The $5\sigma$ discovery region for this channel in the $m_A-\tan\beta$ plane 
is also shown. Evidently
this channel may provide a good opportunity to precisely reconstruct 
the CP-odd Higgs boson mass ($m_A$) for 170 GeV $< m_A < 2m_t$ 
if $\tan\beta$ is close to one.
The impact of SUSY decays on this discovery channel 
might be significant and is under investigation with realistic simulations.

\subsubsection{$H\to hh$}
Observation of this channel would be particularly interesting as information about two different Higgs bosons and their
coupling could be obtained. 
The dominant decay here is the to final state $b\overline{b}b\overline{b}$. However it is not clear how this mode could be triggered
efficiently and there is a very large background from QCD events. The channel $H\to hh\to b\overline{b} \tau^+\tau^-$ would be
triggerable if one tau decayed leptonically. This channel has not been studied.

The decay channel $H\to hh \to \gamma\gamma b\overline{b}$ is triggerable and was studied\cite{froid96} recently. Events were
required to have a pair of isolated photons with $\abseta<2.5$ and $p_T>20$ GeV and two jets with $p_T> 15 (30)$ GeV and
$\abseta<2.5$ at low (high) luminosity. One of the jets was required to be tagged a b-jet and an efficiency of 60 (50) \% assumed
with a rejection of 100 (10)  
against light (charm) jets. The dominant background arises from $\gamma\gamma$ production in
association with light quark jets and is approximately 10 times larger than the $\gamma\gamma b\overline{b}$ background. 
Event rates are very low, for $M_H\sim 230$ GeV and $m_h=70$ GeV  there are about 20 signal events at high luminosity. However the
very small background ($\sim 2$) and the sharp peak in the $\gamma\gamma$ mass distribution should provide convincing evidence of a
signal.

\subsubsection{Other possibilities}

For large masses, the $A$ and $H$ decay almost exclusively to 
$t\overline{t}$. The background in this channel arises from
$t\overline{t}$ production and is very large. A statistically significant 
signal can be extracted provided that the background can be
calibrated \cite{froid96}. For an integrated luminosity 
of $10^5$ pb$^{-1}$ there are about 9000 events 
for $M_A\sim 400$ GeV after, cuts
requiring an isolated lepton (which provides the trigger) and a 
pair of tagged b-quark jets. The $t\overline{t}$ mass resolution is
of order 70 GeV resulting in approximately 100000 background events. 
The rate for $t\overline{t}$ production is well predicted by
perturbative QCD, so it may well be possible to 
convincingly establish an event excess but extraction of a 
mass for $A$  will be very difficult. The mode is most likely to 
be useful as confirmation of a signal seen elsewhere.

\begin{figure}[t]
\epsfxsize=8.5cm
\dofig{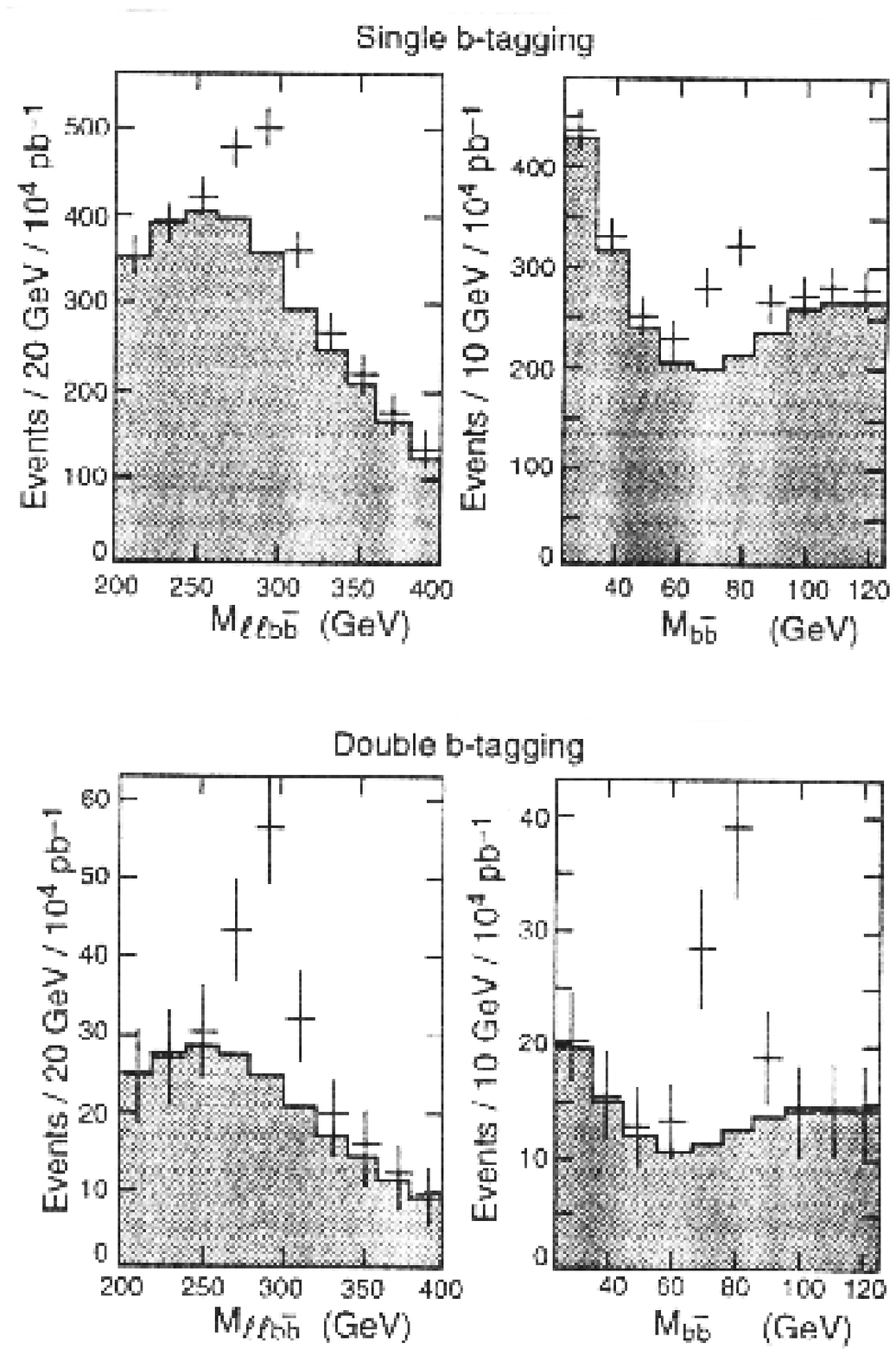}
\caption[]{Reconstructed $\ell\ell b\overline b$ and $b \overline b$
mass distributions from the process $A\to Zh\to\ell\ell b \overline b$. The peaks above
the SM background correspond to the reconstructed $A$ and $h$ from CMS.
\label{cmszh}}
\end{figure}

The decay $A \to  Zh$ affords another place where 
two Higgs bosons might be observed simultaneously.
The leptonic decay of the $Z$ can be used as a trigger. 
The CMS study
requires a pair of electrons (muons)  with $p_T >20 $ (5) which have 
an invariant mass within 6 GeV of the $Z$ mass and
 a pair of jets with $p_T>40$ GeV. One or two 
b-tags are required with an assumed efficiency of 40\% and a rejection of 50 
against light
 quark jets.  The background is dominated by $t\overline{t}$ events.\footnote{A K factor of 1.5 was included in the backgrounds
 shown on figure \ref{cmszh}.} The signal to background ratio is much better than in the case of $WH (\to b\overline{b}$) as can
 be seen in figure
 \ref{cmszh}. A peak is clearly visible in both the $b\overline{b}$ and $\ell\ell b\overline{b}$ mass distributions and a signal
 could be unambiguously seen.

The positive conclusion of this study is confirmed in ref \cite{froid96} where several values of $M_A$ and $m_h$ were simulated and
it was concluded that a $5\sigma$ signal is observable for an integrated luminosity of $3\times 10^{4}$ pb$^{-1}$ for $\tan\beta
<2$ and $150 <M_A< 350$. This study included the background from $Zb\overline{b}$ events which dominate over the $t\overline{t}$
background at smaller values of $m_A$.

\subsubsection{Summary of Supersymmetric Higgs}

One is confident that the following modes will be effective in searching for the MSSM Higgs bosons: $A/H \to \tau^+\tau^-$,
$A/H\to \mu^+\mu^-$, $H\to Z Z^* \to 4\ell$, $h\to \gamma\gamma$, $A\to Zh\to \ell\ell b \overline{b}$, $H\to h h \to
b\overline{b}\gamma\gamma$  and  $t \to b H^+(\to \tau \nu)$
(discussed in the section on the top quark). In addition, the modes $A/H\to t\overline{t}$ and $h\to b\overline{b}$  for $h$
produced in association with a $W$ may provide valuable information. The former set of modes are sufficient for either experiment 
to {\it exclude} the
entire $tan\beta - M_A$ plane at 95\% confidence with $10^5$ pb$^{-1}$.

\begin{figure}[t]
\epsfxsize=8.5cm
\dofig{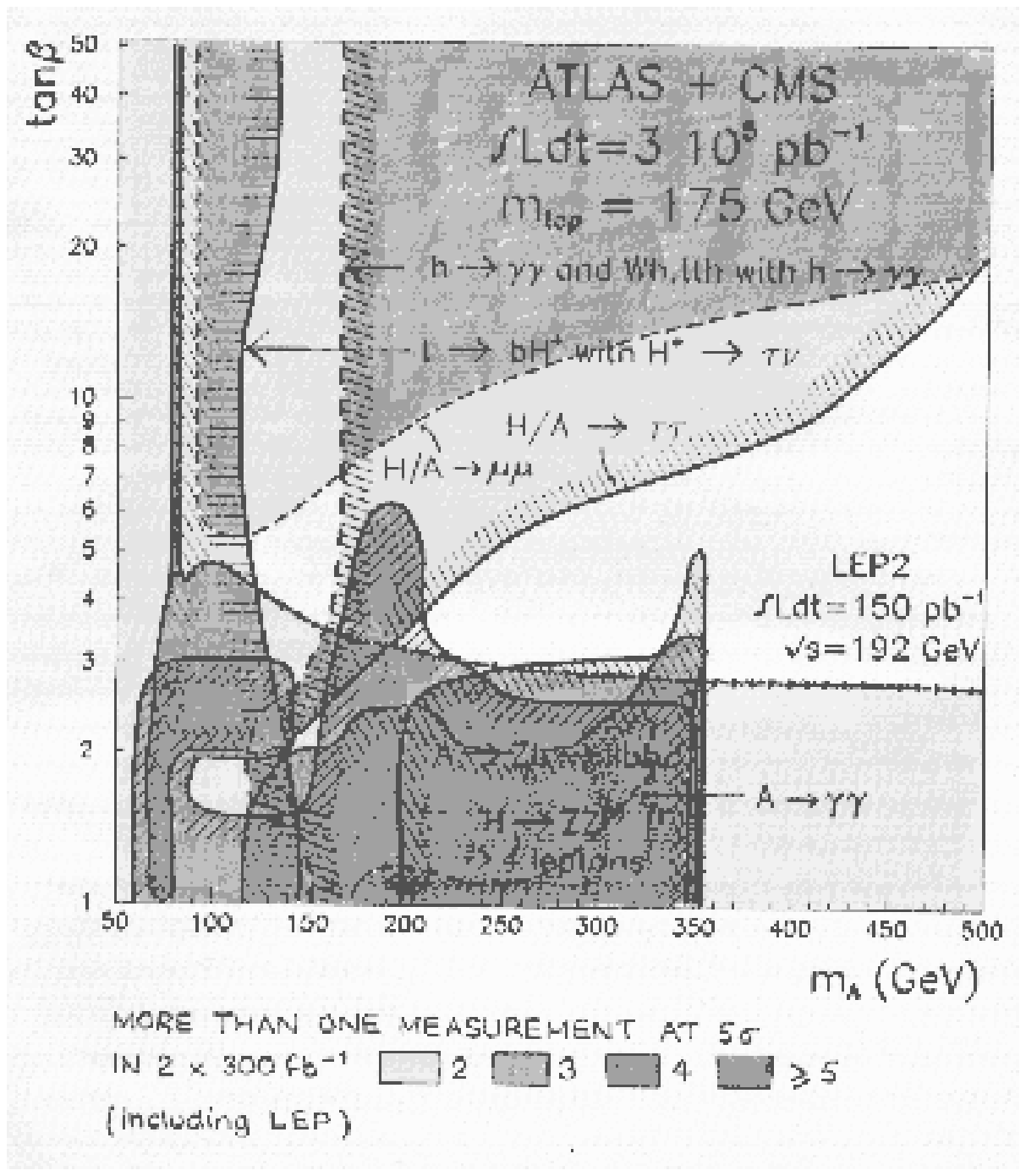}
\caption[]{$5\sigma$ exclusion contours for the various processes used to search for Higgs bosons in the MSSM. 
\label{froidfig}}
\end{figure}

Ensuring a $5\sigma$ discovery over the entire $tan\beta - M_A$ plane 
requires more luminosity. Figure \ref{froidfig}
shows an indication of what 
can be achieved after a few years of running\cite{froid96}.  The entire plane is 
covered using the modes where one has great
confidence. Over a significant fraction of the parameter space at least two 
distinct modes will be visible.  For example, if $h$ is observed at
LEP II and $M_A$ is small the LHC will see the $H^+$ in top quark decay,
$H \to ZZ^*$, and possibly $H/A\to\tau\tau$.
At large values of $M_A$, the decays $h\to \gamma\gamma$, $H\to ZZ^*$, and $A \to Zh$
will provide a third or fourth observation.  If nothing is observed at 
LEPII, then over a significant fraction of
the remaining phase space, $h\to \gamma\gamma$ and $H/A\to \tau\tau$ (and $H/A \to \mu\mu$) 
will be measured.  
%There is  region of
%parameter space at large $M_A$ and moderate $\tan\beta$ where only $h\to \gamma\gamma$ will be available.
The decay of other supersymmetric particles will provide additional sources of $h$.  
Over a significant fraction of SUSY parameter space, there is a substantial branching fraction for 
squarks to decay to $h$. The rate is then such that decay $h\to b\overline{b}$ becomes clearly observable above background
and this channel is the one where $h$ is observed first at LHC. 

\section{Supersymmetry} 

The supersymmetric extension to the standard model has a rich spectrum of 
particles that can be observed at
the LHC. In addition to the extended Higgs sector discussed  above, there are the 
supersymmetric partners of all the quarks, leptons
and gauge bosons of the standard model. If supersymmetry is relevant to the 
electroweak symmetry breaking problem then most of
these  particles  will be  in a mass range that is
observable at LHC \cite{anderson} The  sparticles  with the  largest  production  rates 
at LHC  are those  with strong
interaction  couplings, the  squarks and  gluinos.  Production  rates are very  
large and the  discussion then  must focus on decay
scenarios.

Many  supersymmetric models have  a discrete  symmetry called  R-parity that  ensures that the  lightest  supersymmetric particle is
absolutely stable. This particle must be electrically neutral and might pervade all of the current universe providing a substantial
fraction of the dark matter. This particle  could be the partner of the neutrino  (sneutrino), but in most supersymmetric models it
is one of the four mass eigenstates that are linear  combinations the partners of the $Z$, $\gamma$, and neutral components of the
two Higgs doublets. These  states (in order of  increasing  mass) are denoted  by $\chi_1$,  $\chi_2$, $\chi_3$  and $\chi_4$. The
production rates for these particles are small and their largest  source is the decay of other supersymmetric particles. Since these
so-called  neutralinos have no  electric charge and no  strong  interactions, they have very  small interaction  cross-sections off
regular  matter. The  lightest of  them exits  the  detector carrying  off energy  and leading  to one of  the classic  signals for
supersymmetry at a hadron collider: missing $E_T$.

Heavier neutralinos can decay into lighter ones  via the emission of a 
(real or  virtual) $Z$ boson. The partners of
the $W$ boson ($\chi^{\pm}$) can either be  produced directly or in the decay of other  
supersymmetric particles ($ \eg \tilde{g}\to
q\overline{q} \chi^{\pm}$). The subsequent decay of 
a $\chi^{\pm}$ will give rise to a (real or virtual) $W$ boson  (\eg $\chi^{\pm}
\to W \chi_1$) and hence to an isolated  leptons. Since the gluino is a Majorana  
fermion its decay can lead to either $\ell^+$ or
$\ell^-$. This observation leads to the second characteristic  signature, Events 
with one, two or three isolated leptons in various
charge combinations. The  final state with a pair  of isolated leptons of 
the  same charge is  particularly interesting as standard
model physics (such as the production of a $t \overline{t}$  pair) leads 
to a rate for this that is much below that for an isolated
lepton pair of opposite charge.

The mass spectrum and detailed decay properties of the supersymmetric particles are 
very model dependent making a general
study rather difficult. The situation  is  complicated by the  real  possibility that the  LHC may be a  factory for  supersymmetric
particles; many different ones are  produced at the same time. Early studies of  supersymmetric signals concentrated on a specific
particle and a particular decay  mode demonstrating that cuts  could be made that ensure that the  signal from this decay stands out
above the standard model background. These studies provide a convincing case 
that supersymmetry could be discovered at the LHC. The
next level of work addresses the question  of how the masses and 
couplings of the  particles could be determined and the underlying
theory constrained.
Here one faces the problem that the dominant background for 
supersymmetry is supersymmetry itself. 

So far, direct searches at the Tevatron have excluded the mass range up
to $m_{\gluino} = 230\,$GeV.\cite{tevsusy}  With the main injector and 2~fb$^{-1}$ of
luminosity, sensitivity will extend up to $m_{\gluino} 
\sim 300-400\,$GeV\cite{tev2k},
and if we are lucky we might see something.  One of the great strengths of the
LHC, however, is that it is sensitive to supersymmetry over the whole mass range
over which the theory makes sense (at least as far as electroweak symmetry breaking is concerned), i.e.
$m_{\gluino} \simle 1-1.5\,$TeV.  

\subsection{Squark and Gluino searches}

There are two distinct sources of background for the supersymmetry signatures 
involving jets and missing $E_T$. The first is real physics
processes  involving, for example,  jets produced in  association with 
a  $Z$ boson that then  decays to  $\nu\overline{\nu}$. These
backgrounds are detector independent and irreducible. Secondly,  
backgrounds arise from the mismeasurement of multijet jet final
states due to  imperfections in the  detectors. This  can happen because  of 
poor jet energy resolution which  then allows a jet's
energy to be substantially mismeasured resulting in apparent missing $E_T$, 
or cracks and dead material which cause energy to be lost.  
This background, if it proves to be important, can be
reduced by rejecting events where the missing $E_T$ vector is closely 
aligned with one of the jets.

\begin{figure} 
\epsfxsize=8.5cm
\dofig{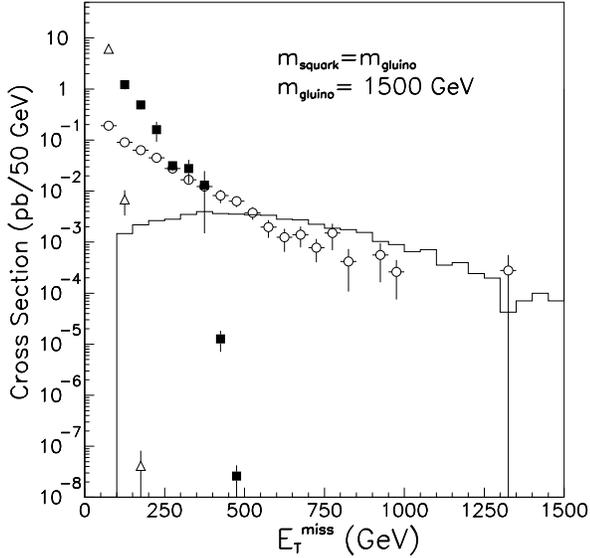}
\caption[]{Missing $E_T$ signature  arising from a  supersymmetry event having at  
least three jets with
$E_T>200$ GeV, a fourth  jet with $E_T>100$ GeV and  transverse sphericity  
$S_T>0.2$. The solid  histogram is the signal, the open
circles are the irreducible background arising from the decay into neutrinos 
of $t$, $W$, $Z$ \etc. The filled squares represent the
reducible background in the unreasonable case where all the energy in the 
region between $\abseta$ of 3.1 and 3.3 is lost. The more
realistic case of degraded resolution is shown as the triangles. Figure from an ATLAS simulation. } 
\label{sys1} \end{figure}

ATLAS conducted a study  of the second  background. A sample of four and  five jet events was produced  using exact matrix element
calculations  interfaced with JETSET 7.4.  This method of  generation is expected to  be more reliable for  events with many widely
separated jets than that from a  showering Monte-Carlo alone. A  parameterization of a GEANT based  study \cite{atlascal}
of the jet response in the
potentially troublesome  region between the  forward and end cap  calorimeters was used. The resulting  in background is far below
that from the irreducible background as is shown in Figure \ref{sys1}

From this figure it can be seen  that the completely unrealistic  case 
where all the energy in  the region $3.1<\abseta <3.3$ is lost still
produces a background that is far below the irreducible background. This study 
confirms ones done for the SSC\cite{sdc} that indicate
that these reducible backgrounds are unimportant. Figure \ref{sys1} also 
shows that at sufficiently large missing $E_T$, the signal
from the decay of squarks and gluinos exceeds that from standard model 
background sources. 

A similar study was  carried out by CMS\cite{cmssquark}. Here the  
MSSM was used as implemented in 
ISASUSY \cite{isasusy}. The  following parameters were chosen:
$M_{\tilde{g}}=1500$ GeV, $m_{\tilde{q}}=1550 $ GeV $\mu=-440$  GeV, $\tan\beta=2$, $m_{\tilde{l}}=300$ GeV. Events were selected
that have a least 4 jets with $E_T>100$ GeV, one of them was  required to have $E_T>400 GeV$ and another to have $E_T>200$ GeV. The
three highest $E_T$ jets have $\abseta <1.5$ and the other has  $\abseta<2.0$, $\ETmiss >600$ Gev, circularity greater than 0.1 and
the invariant mass of the jets and the missing $E_T$ is at least 1500 GeV. For this very massive gluino and squark case, 450 events
survive these cuts for an  integrated luminosity of  $10^5$ pb$^{-1}$. There  are 90 background events,  58 of which arise from the
production of $W$ and $Z$ bosons in association with jets 
so a signal can be clearly established.
Both ATLAS and CMS can discover gluinos up to $m_{\tilde{g}}=1500$ GeV.

\subsubsection{Jets and Leptons}

ATLAS conducted a simulation  of same sign  dilepton signals.  Here the  dominant background  arises from  $t\overline{t}$ events:
$t\overline{t}\to W (\to \ell^+) b W  \overline{b} (\to \ell^+ \nu c)$. The  requirement that both leptons be isolated (less than 12
(5) GeV of additional energy in a cone of  size $\Delta R=0.2 (0.3)$ around the lepton  direction at high (low) luminosity), is very
effective at reducing the  background from bottom decays. Events  were required to have two  isolated leptons with $p_T>20 GeV$ and
$\abseta <2.5$, four jets with  $E_T>70 (110)$ GeV( at least one  of these has $E_T>110(150)$  GeV) and $\ETmiss >120 (150)$ GeV at
low (high) luminosity. For an integrated luminosity of $10^5$  pb$^{-1}$ and $m_{\tilde{g}}\sim m_{\tilde{q}}$ there are about 14000
(120) signal events over a background of 500 (70) for $m_{\tilde{g}}=300 (1500)$ GeV. 
The results of this study can be converted into
a reach  in the  MSSM. For  most  values of   $\tan\beta$ and  $\mu$, and  for   $m_{\tilde{g}}\sim  m_{\tilde{q}}$  
 ($m_{\tilde{g}}\sim
2m_{\tilde{q}}$) [$2m_{\tilde{g}}\sim m_{\tilde{q}}$], gluino masses up to 1800 (2600) [1400] GeV can be probed in this  channel.

If squark production is dominant, there will be an asymmetry in  the signs of the dilepton pairs that arises because the beams are
protons which  contain more up  than down type  quarks. This  asymmetry is  
$$A=\frac{\sigma (++)-\sigma  (--)}{\sigma (++)+\sigma(--)+{\rm background}}$$ 
The asymmetry is very small if  $m_{squark}=2m_{gluino}$ but it rises to  $A\sim 0.2$ for $m_{squark}=m_{gluino}/2$ and for
this value could be  measured with a precision of  $\delta A=0.05$ up to  squark masses of 750 GeV.  This quantity is an example of
ones that will be used to pin down the details of the supersymmetry spectrum after a signal has been observed.

CMS have also investigated muon(s)$+$jets$+\met$ signatures for 
supersymmetry\cite{cmssquarklep}.
Channels with a single muon, two muons of the same or of any sign, two isolated
muons, and three muons were investigated.
These channels are found to allow the observation of a gluino signal up to
$m_{gluino}\approx 1.5\,$~TeV with $10^5\,{\rm pb}^{-1}$.

\subsection{Charginos and Neutralinos}

The pair production of charginos and neutralinos will result in final 
states with three isolated leptons from the decay chains $\chi^{\pm}\to
\ell\nu \chi_1$ and  $\chi_2\to\chi_1  \ell^+\ell^-$. After isolation  requirements on the leptons, the  dominant background is from
$WZ$ final states. This final state has been used at the Tevatron. No 
signal was observed allowing a cross section limit to be set\cite{tevsusy}.

ATLAS used the MSSM to investigate the  utility of this mode at LHC. Three isolated  leptons with $\abseta<2.5$ were required, two
of which have $p_T>20$ GeV and the third has $p_T>10 $ GeV.  Events were rejected if they are had a lepton pair consistent with the
decay of a $Z$  (reconstructed mass within 10 GeV of  the $Z$ mass). This cut  did not reduce the signal  because, over the parameter
space searched, $m_{\chi_2}-m_{\chi_1} <80$ GeV. MSSM parameters  $\tan\beta=2$ and 20, $\mu=-m_{gluino}$, $m_{squark}=2m_{gluino}$
and  $m_{squark}=m_{gluino}+20  $ GeV for   $m_{gluino}=200,300,400,500$ and 600  GeV were used.  A jet veto to  reduce further the
$t\overline{t}$ background was used (no jets with $p_T>25 $ GeV and $\abseta <3$) although this cut may not be needed (check this).
At low luminosity $10$ fb$^{-1}$ (the jet veto is questionable at high luminosity), there is a statistically significant signal up
to gluino masses of 600 (400) GeV at the smaller (larger)value of $\tan\beta$.

\subsection{Sleptons} 

The partners of the leptons are the most  difficult supersymmetric particles to observe at a hadron collider.
Their production rate is very small as it  is dominated by the Drell-Yan process  $q\overline{q} \to \tilde{\ell^+}\tilde{\ell^-}$,
unless sleptons are produced in the decays of strongly interacting sparticles.
The slepton 
decays will  produce final states of  opposite sign lepton  pairs and missing  $E_T$. Backgrounds  notably from $t\overline{t}$
final states are  very large. All hope of  extracting a signal  relies on the  efficient use of a jet veto.  Simulations of slepton
signals have not yet been carried out with the ATLAS or CMS detectors. However a ``toy simulation'' with some degree of credibility
indicates that it might be possible to extract a signal\cite{paigeslept}.
Events were selected requiring that there be a pair of isolated leptons of the same flavor and 
opposite charge and $p_T> 20$ GeV.
At least 100 GeV of missing $E_T$ was required and events were vetoed if the was a jet with $p_T>25$ GeV and $\abseta<3$. The
missing $E_T$ vector and the transverse momentum sector of the dilepton system were more than $160^o$ apart in azimuth. The
dominant background is from $t\overline{t}$ and $W^+W^-$ events. Slepton masses up to about 300 GeV
are observable with $10$ fb $^{-1}$ of integrated luminosity. The signal is not obscured 
by other SUSY decays.
This study is very encouraging, a more detailed simulation is
required to confirm it. 
Such investigations are now in progress in CMS, including the question of
separating the slepton signal from the backgrounds arising from the copious
production of other SUSY states.  

\subsection{Which SUSY?} 

From the studies described above and others one has give absolute  confidence that the LHC can discover supersymmetry if
it is kinematically  accessible. The more difficult question of 
how well masses and  branching ratios can be  measured has recently begun to be studied. The proliferation of
models makes a  systematic  approach difficult.  In a seminal  work, 
Paige \etal\ have  investigated the  dependence of the signals
discussed above and many others  upon the parameters in the  minimal supergravity model (SUGRA)\cite{paige}.  
This model has the advantage that
rather few parameters specify it completely. The model is assumed to unify at some high scale where a common gaugino mass $m_{1/2}$
is defined. All scalar  particles are assumed to have a common  mass $m_0$ at this scale. Three  other parameters then fully specify
the model: $\tan\beta$, a  variable $A$ with dimension of mass  that affects mainly the splitting  between the partners of the left
and right handed top  quark, and the sign of $\mu$. Final states involving leptons, jets and missing $E_T$ were investigated to
determine sensitivity to the parameters that an LHC experiment may have. 
One result of this study is that $b-$quark tagging might be
an important tool in disentangling the parameter space as the b-quark multiplicity is a useful quantity to measure.

\begin{figure}
\epsfxsize=8.5cm
\dofig{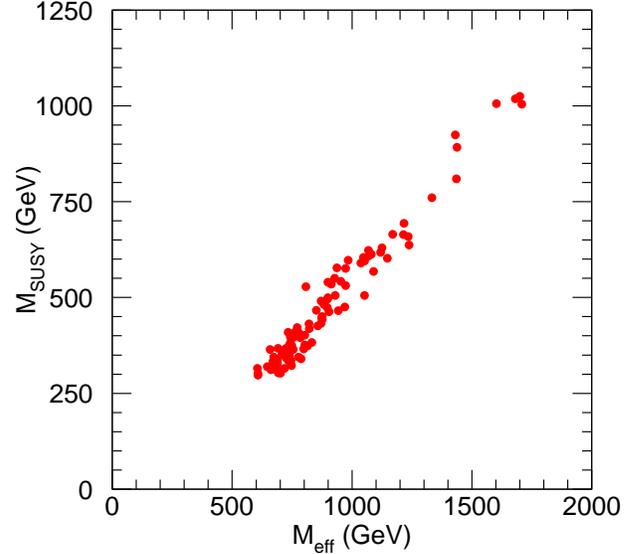}
\caption[]{The correlation between the peak in the $M_4$ distribution, $M_eff$ and $M_{susy}$ being the 
smaller of the gluino and average of the up, down charm and strange squark masses.
\label{susycorr}}
\end{figure}

A study \cite{susysnow} has attempted to address the issue of how well the parameters in a SUGRA model could be determined.
The proposed strategy is as follows. One first searches for an excess of events over background by using several variables.
For example, events are selected which have at least 4 jets one of which has $E_t>100$ GeV and the others have $E_T>50$.
An additional requirement of $\ETmiss >100$ GeV and sphericity $S>2$ is made, and the event rate is plotted against $M_4$ defined
as the scalar sum of the $E_T$ of the four jets and $\ETmiss$ \cite{toypaige}. This curve has a peak in the region where the signal
to background ratio is large and there is a strong
correlation between the position of the peak and the smallest of the gluino and (up, down, strange and charm) squark masses as is
shown in Figure \ref{susycorr}. This correlation can then be exploited to determine the overall mass scale for the strongly
interacting superparticles with an accuracy of order 10\%. 

Having determined the scale, more detailed measurements are then performed. For this purpose a particular point in the parameter
space was selected for simulation. The mass spectrum is as follows: Gluino $m_{\tilde{g}}=298$ GeV 
 $m_{\tilde{q_r}}=312$ GeV, $m_{\tilde{q_l}}=317$ GeV
 $m_{\tilde{t_1}}=263$ GeV,
$m_{\tilde{t_2}}=329$ GeV \\$m_{\tilde{b_1}}=278$ GeV, $m_{\tilde{b_2}}=314$ GeV
Sleptons $m_{\tilde{e_l}}=215$ GeV,  $m_{\tilde{e_r}}=206$ GeV,
  Neutralinos $m_{\chi_1}=44$ GeV, $m_{\chi_2}=98$ GeV,
 $m_{\chi_3}=257$ GeV, $m_{\chi_4}=273$ GeV
Charginos $m_{\tilde{\chi^+_1}}=96$ GeV, $m_{\tilde{\chi^+_2}}=272$ GeV
 Higgs $m_{{h}}=68$ GeV, $m_{{H}}=378$ GeV,
$m_{{A}}=371$ GeV, $m_{{H^+}}=378$ GeV.  

At this point the total production rate for gluino pairs is very large, and many other
supersymmetric particles are produced in the decay of gluinos. Of particular significance is $\chi_2$ which decays to $\chi_1
e^+e^-$ and $\chi_1\mu^+\mu^-$  with a combined branching ratio of 32\%. The position of the end point of this spectrum
determines the mass difference $m_{\chi_2}-m_{\chi_1}$ \cite{jesper}.
 Backgrounds are negligible if the events are required to have two such
dilepton pairs, which can arise from the pair production of gluinos with each decaying to $b\tilde{b}(\to \chi_2 (\to
\chi_1\ell^+\ell^-))$ which has a combined branching ratio of 24\%. The event rate is so large that the statistical error in the
determination of the mass difference is very small and the total error will be dominated by systematic effects. The enormous number
of $Z\to\ell^+\ell^-$ decays can be used to calibrate, and an error of better than 50 MeV on 
$m_{\chi_2}-m_{\chi_1}$ is achievable\footnote{Recall that the current error on $M_W$ from CDF/D0 \cite{tevwmass} comes from an
analysis involving $\ETmiss$ has far fewer events and has an error of order 150 MeV}. In the context of the model, this measurement
constrains $M_{1/2}$ with an error of order 0.1\%. By comparing event rates for samples with one or two dilepton pairs, the
branching ratio $\chi_2\to\chi_1 \mu^+\mu^-$ can be measured. 

\begin{figure}[t]
\epsfxsize=8.5cm
\dofig{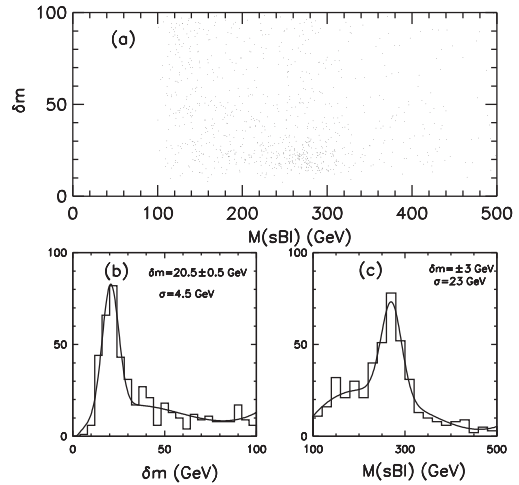}
\caption[]{The reconstruction of gluino and sbottom decays from the decay chain $\tilde{g}\to \chi_2 (\to\chi_1 \ell^+\ell^-)
\tilde{b}$. Events are selected near the end point of the $\ell^-\ell^+$  
mass distribution and the momentum of $\chi_2$ reconstructed. Two $b-$jets are then required and the mass of $b+\chi_2$
($=m_{\tilde{b}}$ and the mass difference $\delta m=m_{bb\chi_2}-m_{b\chi_2}$ are computed. The scatterplot in these two variables
and the projections are shown. 
\label{yaofig}}
\end{figure}

The small mass difference between the gluino and the sbottom can also be exploited to reconstruct a the masses of these particles
\cite{weiming}. Here a partial reconstruction technique is used. Events are selected where the dilepton invariant mass is close
to its maximum value. In the rest frame of $\chi_2$, $\chi_1$ is then forced to be at rest. The momentum of $\chi_2$ in the
laboratory frame is then related to the momentum of the $\ell^+\ell^-$ pair by
$p_{\chi_2}=(1+m_{\chi_1}/m_{\ell^+\ell^-})p_{\ell^+ \ell^-}$. $\chi_2$ can then be combined with an additional $b-jet$ to
reconstruct the $tilde{b}$ mass. An additional $b_jet$ can then be added to reconstruct the $\tilde{g}$ mass. 
Figure 
\ref{yaofig} shows the scatterplot on these two invariant masses together with a projection onto $m_{\tilde{b}}$ and
$\delta m=m_{tilde{g}}-m_{\tilde{b}}$. Peaks can clearly be seen above the combinatoric background.
This method can be used to determine $m_{\tilde{g}}$ and $m_{\tilde{b}}$. The values depend on the assumed value of $m_{\chi_1}$:
$m_{\tilde{b}}=m_{\tilde{b}}^{true}+1.5(m_{\chi_1}^{assumed}-m_{\chi_1}^{true}) \pm 3 $GeV and  $m_{\tilde{g}}-m_{\tilde{b}}=
m_{\tilde{g}}^{true}-m_{\tilde{b}}^{true} \pm 0.5$ GeV. 

Once several quantities have been measured, one will attempt to constrain the parameters of the SUSY model by performing a global
fit much as the standard model is tested at LEP \cite{LEP}. To get and indication of how well this might work, many choices of
parameters within the SUGRA model were made and those that resulted in masses within the expected error were retained
\cite{susysnow}. Measurements of $m_h$, $m_{\chi_2}-m_{\chi_1}$ and $m_{\tilde{g}}-m_{\tilde{b}}$ with errors of $\pm 5$ GeV,
 $\pm 0.50 $GeV$ (10\sigma)$ 
and $\pm 3$ GeV $(1.5\sigma)$  respectively result in the constraints $\delta m_{1/2} =1.5$ GeV, $\delta m_0=15$ GeV and $\delta
\tan\beta=0.1$. It is clear from this example that precise measurements of SUSY parameters will be made at LHC if supersymmetric
particles exist.

Other supersymmetric models  such as the recently popular models  where supersymmetry is broken  at a rather low scale\cite{dine},
 can produce
signals different from SUGRA models. In particular $\chi_1$ may be unstable and may decay to $\gamma+\tilde{G}$, reducing the
missing $E_T$ rate ($\tilde{G}$ exits unobserved) but providing {\bf every} supersymmetry event with an additional pair of isolated
photons. We should hope that these models are correct 
as this signal is trivial to observe at LHC.

\section{Strong Dynamics}

\subsection{Strongly interacting $W$'s}

The couplings of longitudinally polarized gauge bosons to each other  are fixed at low energy
by the nature of the spontaneously broken electro-weak symmetry and are independent of the details of the
breaking mechanism. Scattering amplitudes calculated from these couplings will violate 
unitarity at center of mass energies
of the $WW$ system  around 1.5 TeV. New physics must enter to cure this problem.
In the minimal standard model and its supersymmetric version, the cure arises from the perturbative couplings
of the Higgs bosons.   
If no Higgs-like particle exists, then new  non-perturbative dynamics must enter in the 
scattering amplitudes for $WW$, $WZ$ and $ZZ$ scattering at high energy.
Therefore if no new physics shows up
at lower mass scales one must be able to probe $W_L W_L$ scattering at 
$\sqrt{\hat s} \sim 1 $ TeV. 

Various models exist that can be used as benchmarks for this physics\cite{chanowitz}. The basic signal in all of them is an excess of events over
that predicted by the standard model for gauge boson pairs of large invariant mass. In certain models resonant structure can be
seen (an example of this is given in the next subsection). 
In the standard model, the   $W^+W^+$ final state is
the only one where there is no process $q\overline{q}\to WW$ and is therefore 
expected to have a much smaller
background than, for example, the $ZZ$ or $W^+W^-$ final state. Background 
is present at a smaller level from
$q\overline{q}\to W q\overline{q}$ proceeding either by 
gluon exchange or via an order $\alpha^2$ electroweak process and from
the final state $Wt\overline{t}$. There is a background from $WZ$ if one lepton is lost.
There is negligible background from charge misidentification in either ATLAS or CMS.

ATLAS \cite{atlaswwnote} conducted a parton model study of the 
signal and background in this channel. Events were selected that have two leptons of
the same sign with $p_T>25$ GeV and $\abseta <2.5$. If a third 
lepton was present that, in combination with one of the other two,
was consistent with the decay of a $Z$ (mass within 15 GeV of the $Z$ 
mass), the event was rejected. This cut is needed to
eliminate the background from $WZ$ and $ZZ$ final states. In addition the two leptons
are required to have invariant mass above 100 GeV,
to  have transverse momenta within 80 GeV of each other and 
to be separated in $\phi$ by at least $\pi/2$. At this stage, there are
there are $\sim 1700$ standard model events for a luminosity of 
$10^5$ pb $^{-1}$. Of these events roughly 50\% are from
$WZ$ and $ZZ$ final states and 30\% from $Wt\overline{t}$.
There are of order 40 signal events depending upon the model used for 
the strongly coupled gauge boson
sector.  Additional cuts are needed to reduce the background. A 
jet veto requiring no jets with $p_T> 40$ GeV and $\abseta <2$
is effective against the $Wt\overline{t}$ final state. The requirement 
of two forward jet tags each with $15<p_T<130$ GeV
and $\abseta >3$ reduces the $WW$, $ZZ$ and $WZ$ background.

\begin{figure}
\epsfxsize=8.5cm
\dofig{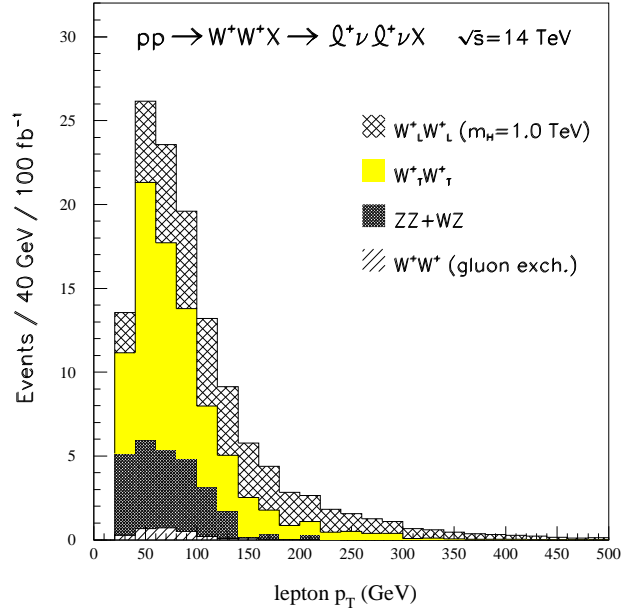}
\caption[]{The $p_T$ spectrum for same sign dileptons in the search for a strongly coupled $WW$ sector as simulated by ATLAS.
 The signal corresponds to a 1
TeV Higgs boson.}
\label{atlasww}
\end{figure}

The remaining background of 40 events is dominated by the 
$q\overline{q}\to W q\overline{q}$ processes. The signal rates vary between 40 and 15 events depending upon the model. The largest
rate arises from a model where the $WW$ scattering amplitude, which is known at small values of $\sqrt{s}$ from low energy theorems
is extrapolated until it saturates unitarity and its growth is then cut off.  A model assuming that the dynamics of $WW$ scattering
is similar to that of $\pi\pi$ scattering in QCD generates approximately 25 signal events.  The case of a 1 TeV standard model
higgs boson is shown in Fig.~\ref{atlasww}. It can be seen that the signal and background have the same shape and therefore the
establishment of a signal requires confidence in the expected level of the background. The experiment is very difficult, but at
full luminosity, a signal might be extracted by comparing 
the rate for $W^+W^+$ with those for $WZ$, $W^+W^-$, and $ZZ$ final states.

A similar study in CMS of the $W^+W^+$ final state leads to similar conclusion \cite{smith96}. 
Jet tagging (vetoing) in the forward (central)
region is essential to extract a signal.

\subsection{Technicolor}

Many models of strong electroweak symmetry breaking (technicolor,
topcolor-assisted technicolor, BESS \cite{bess}) predict resonances which decay 
into vector bosons (or their longitudinal components).  These signals are very
striking since they are produced with large cross sections and may be observed
in the leptonic decay modes of the $W$ and $Z$ where the backgrounds are very
small.  

\begin{figure}
\epsfxsize=8.5cm
\dofig{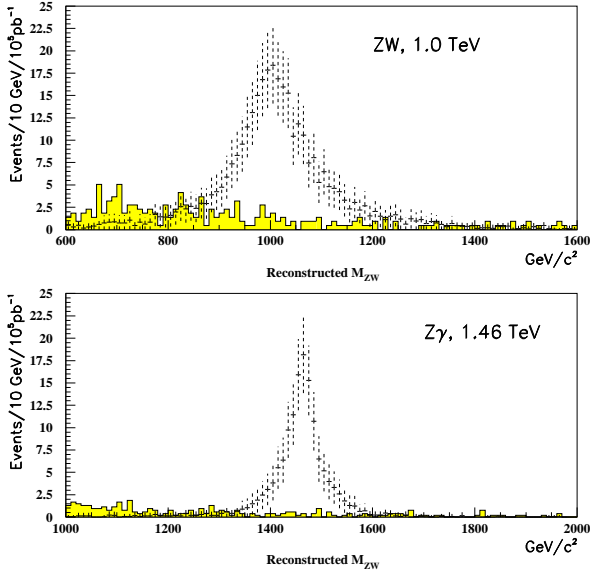}
\caption[]{Reconstructed masses for high-mass resonances decaying
into gauge boson pairs a simulated by ATLAS: $(a)$ $\rho_T$ of mass 1.0~TeV decaying into
$WZ$ and subsequently into 3 leptons; and 
$(b)$ $\omega_T$ of mass 1.46~TeV decaying into
$Z\gamma$ with $Z \to 2$~leptons.}
\label{atlastechni}
\end{figure}

ATLAS have studied a techni-rho, $\rho_T \to WZ$, with
$W\to\ell\nu$, $Z\to\ell\ell$, for $m_{\rho_T} = 1.0$~TeV and also a
techni-omega, $\omega_T \to Z\gamma$, with
$Z\to\ell\ell$, for $m_{\omega_T} = 1.46$~TeV.  The backgrounds due to
$t\overline t$ and continuum vector-boson pair production are small as can be
seen in Fig.\ref{atlastechni}.  

More challenging are the possible decays into non-leptonic modes such as
$\rho_T \to W (\ell\nu) \pi_T (b\overline b)$, which has a signature like
associated $WH$ production with $H \to b\overline b$;
$\eta_T \to t \overline t$, for which the signature is a resonance in
the $t \overline t$ invariant mass; and $\rho_{T8} \to$~jet~jet, 
for which the signature is a resonance in
the dijet invariant mass distribution.

\subsection{Compositeness}

\begin{figure}
\epsfxsize=8.5cm
\dofig{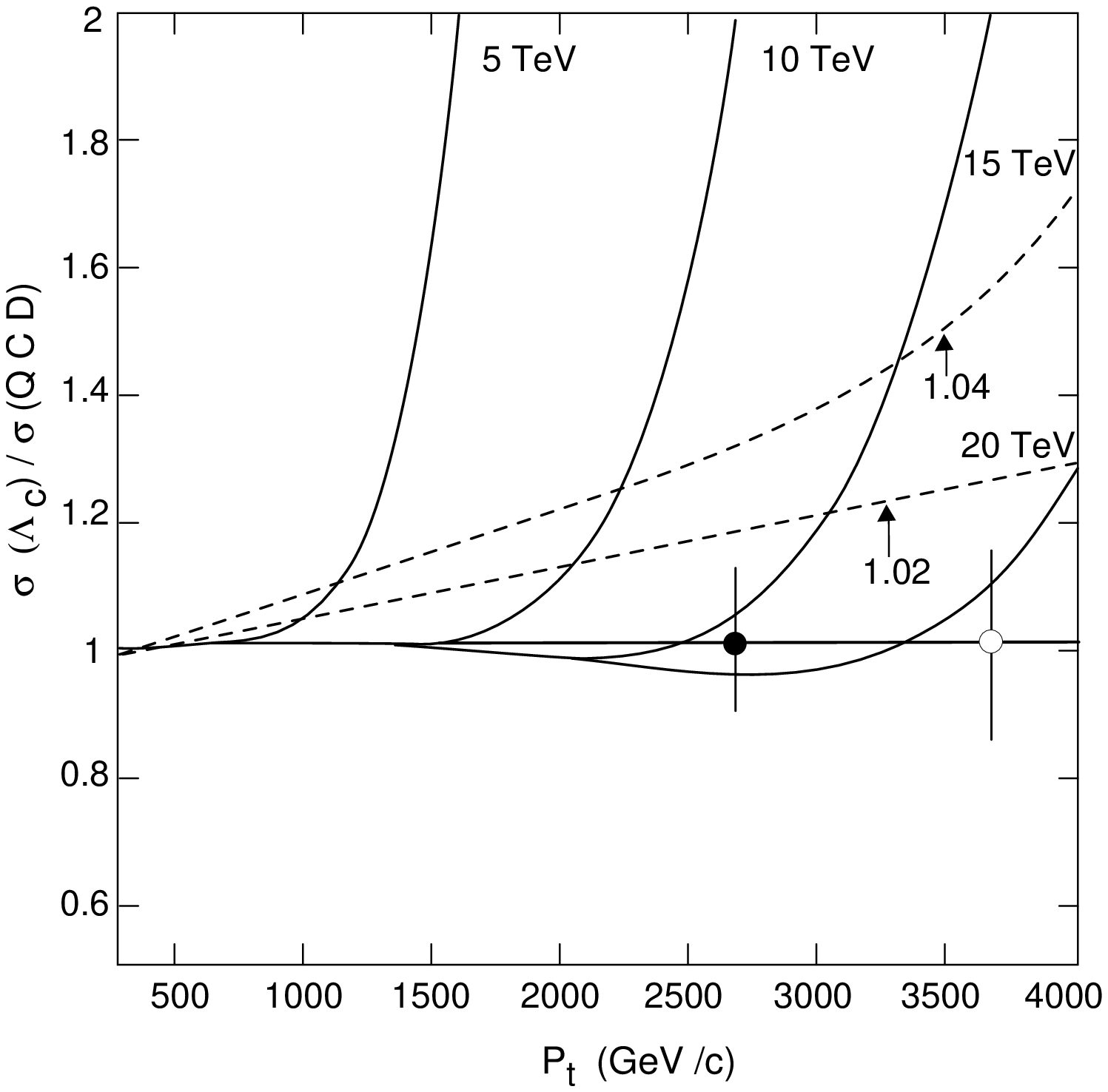}
\label{atlas_compos}
\caption[]{ Deviation from QCD for various values of the compositeness scale $\Lambda$. The error bars correspond to statical
sensitivities at 100 fb$^{-1}$ (open circle) and 10 fb$^{-1}$. The dotted lines refer to the errors induced by possible
nonlinearities in the ATLAS calorimeter.}
\end{figure}

There is no {\it a priori} reason for quarks to be elementary. If they have substructure it will be revealed in the deviations
of the jet cross-section from that predicted by QCD. The deviation is parameterized by an interaction of the form
$4\pi q\gamma^\mu\overline{q}q\gamma^\mu\overline{q}/\Lambda^2$, which has a scale $\Lambda$. 
This is regarded as an effective interaction which is valid only 
for energies less than $\Lambda$.
The ATLAS collaboration has investigated the possibilities for searching for structure in the jet cross-section at high $p_T$.
Figure~\ref{atlas_compos} shows the normalized jet 
cross section $d\sigma/dp_Td\eta$ at $\eta=0$,. The rate is shown as a function of 
$p_T$ for various values of $\Lambda$ and is normalized to 
the value expected from QCD. The error bars at two values of $p_T$
indicate the size of the statistical error to be expected at that value for 
luminosities of $10^4$ and $10^5$ pb$^{-1}$.
It can be seen  that the LHC at full luminosity will be able to probe up to $\Lambda=20$ TeV if the systematic errors are smaller
than the statistical ones. Systematic effects are of two types; theoretical uncertainties in calculating the QCD rates and detector
effects. The former are dependent upon an accurate knowledge of the structure functions in the $x$ range of interest and upon
higher order QCD corrections to the jet cross-sections.  Uncertainties from 
these sources can be expected at the 10\% level.

Experimental effects are of two types. Mismeasurement due to resolution and nonlinearities in the detector response. The former are
at the 20\% level; the latter can be more serious and can induce changes in the apparent shape of the jet cross-section. A
non-linearity at the 4\% level will fake a compositeness signal corresponding to $\Lambda\sim 15 $ TeV. Other distributions, such as
the angular distribution of the jets in a dijet event selected so that the dijet pair has a very large mass, may be
less sensitive to the non-linearites. 

A better reach in $\Lambda$ may be obtained from Drell-Yan dilepton final states, if leptons
are also composite. 

\section{New Gauge Bosons}

A generic prediction of superstring theories is the 
existence of additional $U(1)$ gauge groups.  There is thus
motivation to search for additional $W^\prime$ and
$Z^\prime$ bosons.  
The current Tevatron
limit is 720~GeV for $W^\prime$ (D\O)\cite{tevzprime}.

ATLAS have studied the sensitivity to a new neutral $Z^\prime$ boson
in $e^+e^-$, $\mu\mu$ and jet-jet final states, for
various masses and couplings\cite{atlaszprime}.
It is assumed that $\Gamma_{Z^\prime}\propto m_{Z^\prime}$.
They find the best sensitivity in the $ee$ mode, in which 
signals could be seen up to $m_{Z^\prime}=5$~TeV for standard-model couplings. 
The other final states would provide important information on the $Z^\prime$
couplings. 
The pseudorapidity coverage over which lepton identification and measurement
can be carried out is important for $Z^\prime$ searches: should a signal
be observed, the forward-backward asymmetry of the charged leptons would
provide important information on its nature.  ATLAS found that reducing the
lepton coverage from $|\eta|\leq 2.5$ to $|\eta|\leq 1.2$ roughly halved the
observed asymmetries and prevented discrimination between two particular
$Z^\prime$ models which they investigated.

ATLAS also investigated their sensitivity to a new charged boson
$W^\prime$ decaying into $e\nu$.  The signal is structure in the transverse
mass distribution at masses much greater than $m_W$.  Figure~17 %\ref{atlaswprime}
shows the signal for a 4~TeV $W^\prime$.  They conclude that with
$10^5\,{\rm pb}^{-1}$ one would be sensitive to $m_{W^\prime}=6$~TeV
and that the mass could be measured to 50--100~GeV.

\begin{figure}
\epsfxsize=8.5cm
%\dofig{paige.eps}
\dofig{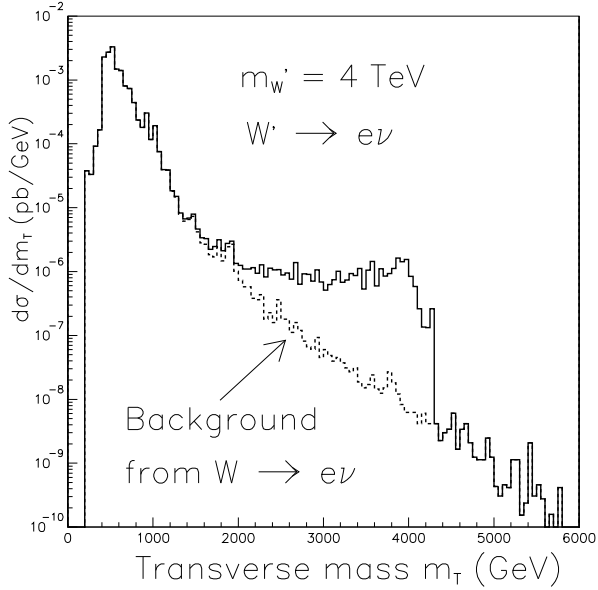}
\label{atlaswprime}
\caption[]{Expected electron-neutrino transverse mass distribution in ATLAS
for $W^\prime \to e \nu$ decays with $m_{W^\prime}=4$~TeV above the
dominant background from $W \to e \nu$ decays.}
\end{figure}

\section{Anomalous Gauge-Boson Couplings}

The trilinear $WWV$ and $Z\gamma V$ couplings ($V=Z, \gamma$) may be probed at
hadron colliders using diboson final states.  
Following the usual notation,
the CP-conserving $WWV$ anomalous couplings are parameterized in terms of 
$\Delta\kappa_V$ and $\lambda_V$,
where $\kappa_V=1$ and $\lambda_V=0$ in the Standard Model for $V=Z, \gamma$.
In general, we would expect anomalous couplings of order $m_W^2/\Lambda^2$ if
$\Lambda$ is the scale for new physics, so if $\Lambda\sim 1\,$TeV then
$\Delta\kappa_V,\lambda_V\sim 0.01$.  
The $Z\gamma V$ anomalous couplings are parameterized in terms of
$h_3^V$ and $h_4^V$, where $h_3^V=h_4^V=0$ in the Standard Model and 
deviations are expected to be ${\cal O}(m_W^4/\Lambda^4)$.   

To maintain
unitarity, the observed anomalous couplings must be modified by a form 
factor; so (for example) 
\begin{equation}
\Delta\kappa_V(q^2) = {\Delta\kappa_V^0 \over (1 + q^2/\Lambda_{FF}^2)^n}
\end{equation}
where $\Lambda_{FF}$ is the form factor scale and $n=2$ for $\Delta\kappa,
\lambda$ and $n=3,4$ for $h_3^V,h_4^V$.  

The ATLAS collaboration have studied\cite{fouchez} their sensitivity 
to anomalous couplings in the $W\gamma$ and $WZ$ modes;
the $W^+W^-$ signal is swamped by $t\overline t$ background. A 
form factor scale $\Lambda_{FF}=10\,$TeV was used.
For the $W\gamma$ final state, events were assumed to be triggered
using a high-$p_T$ lepton plus high-$p_T$ photon candidate.  The background
includes contributions from events with a real lepton and a real photon (e.g.
$b\overline b \gamma$, $t \overline t \gamma$, and $Z\gamma$); a fake lepton 
but a real photon (e.g. $\gamma+{\rm jet}$); and a fake photon with a real
lepton (e.g. $W+{\rm jet}$, $b\overline b$, and $t\overline t$).  
Rejection factors of $10^4$ against jets faking photons and 
$10^5$ against jets faking electrons  were assumed.  
To reduce backgrounds, events
were selected with 
$p_T^\gamma > 100\,{\rm GeV/c}$, $p_T^\ell > 40\,{\rm GeV/c}$,
and $|\eta^\ell|<2.5$.  Events with jets were also vetoed, to further reduce 
backgrounds and to lessen the importance of higher-order QCD corrections.
In an integrated luminosity of $10^5\,{\rm pb}^{-1}$, 7500 events remain, with a
signal to background ratio of 3:1.  The $p_T^\gamma$ distribution is then fitted
in the region where the standard model prediction is 15 events (above about 
$600\,{\rm GeV/c}$), yielding limits of $|\Delta\kappa_\gamma| < 0.04$ and
$|\lambda_\gamma| < 0.0025$ (95\% C.L.).

Similar techniques were used for the $WZ$ state.  The trigger was three
high-$p_T$ leptons, and the backgrounds are from $Zb\overline b$, $Z+{\rm jet}$,
$b \overline b$ and $t\overline t$ processes.  Events were selected with
$p_T^\ell > 25\,{\rm GeV/c}$, $|\eta^\ell|<2.5$, 
$|m_{\ell_1 \ell_2} - m_Z|< 10\,
{\rm GeV/c}^2$, and $m_T(\ell^3, E_T^{miss}) > 40\,{\rm GeV/c}^2$; a jet veto
was also imposed.  In $10^5\,{\rm pb}^{-1}$, 4000
events then remain, with a
signal to background ratio of 2:1.  The $p_T^Z$ distribution is again fitted
in the region where the standard model prediction is 15 events (above about
$380\,{\rm GeV/c}$), yielding limits of $|\Delta\kappa_Z| < 0.07$ and
$|\lambda_Z| < 0.005$ (95\% C.L.).

\begin{figure}[t]
\vskip 12cm
\includegraphics{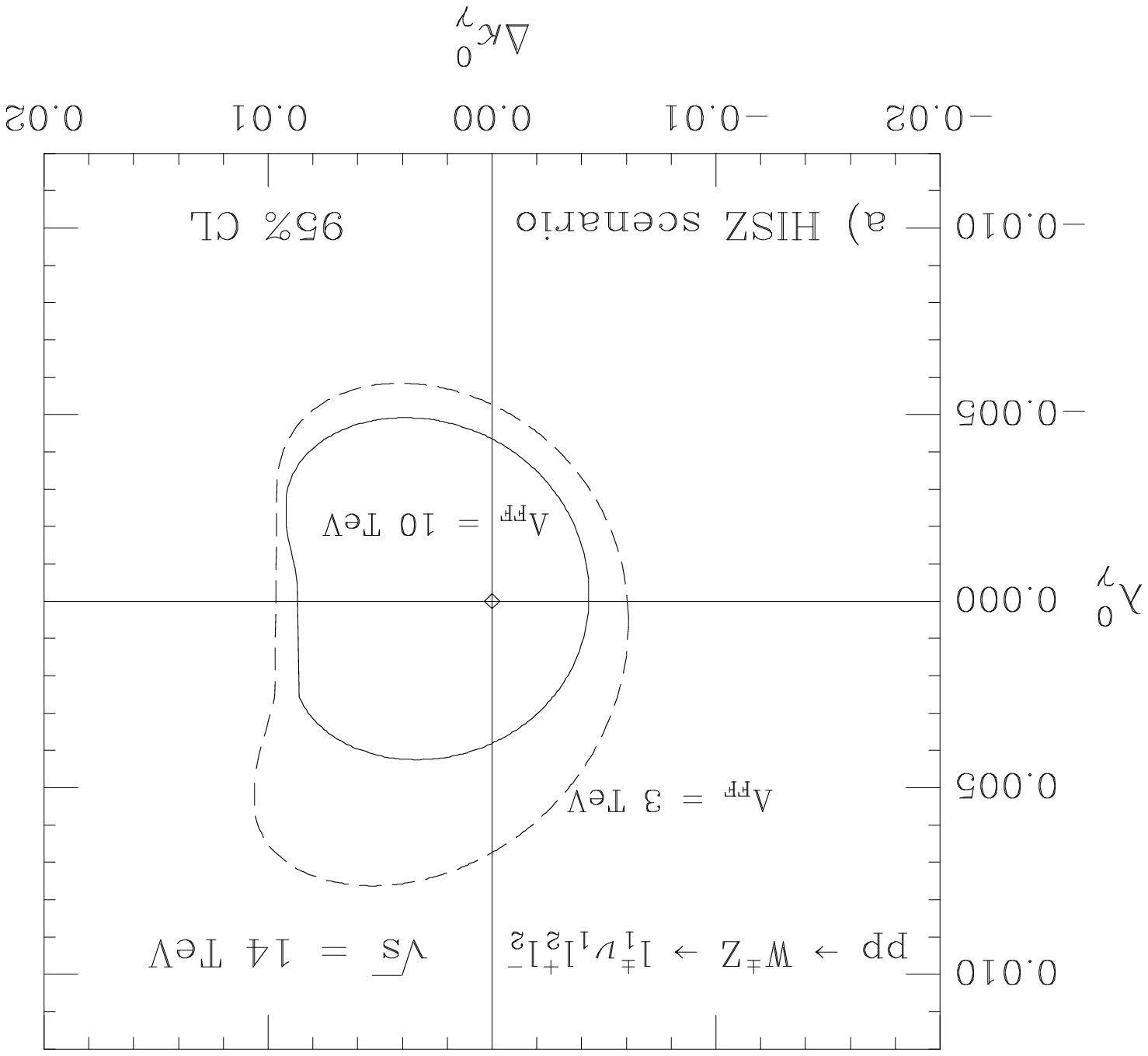}
\includegraphics{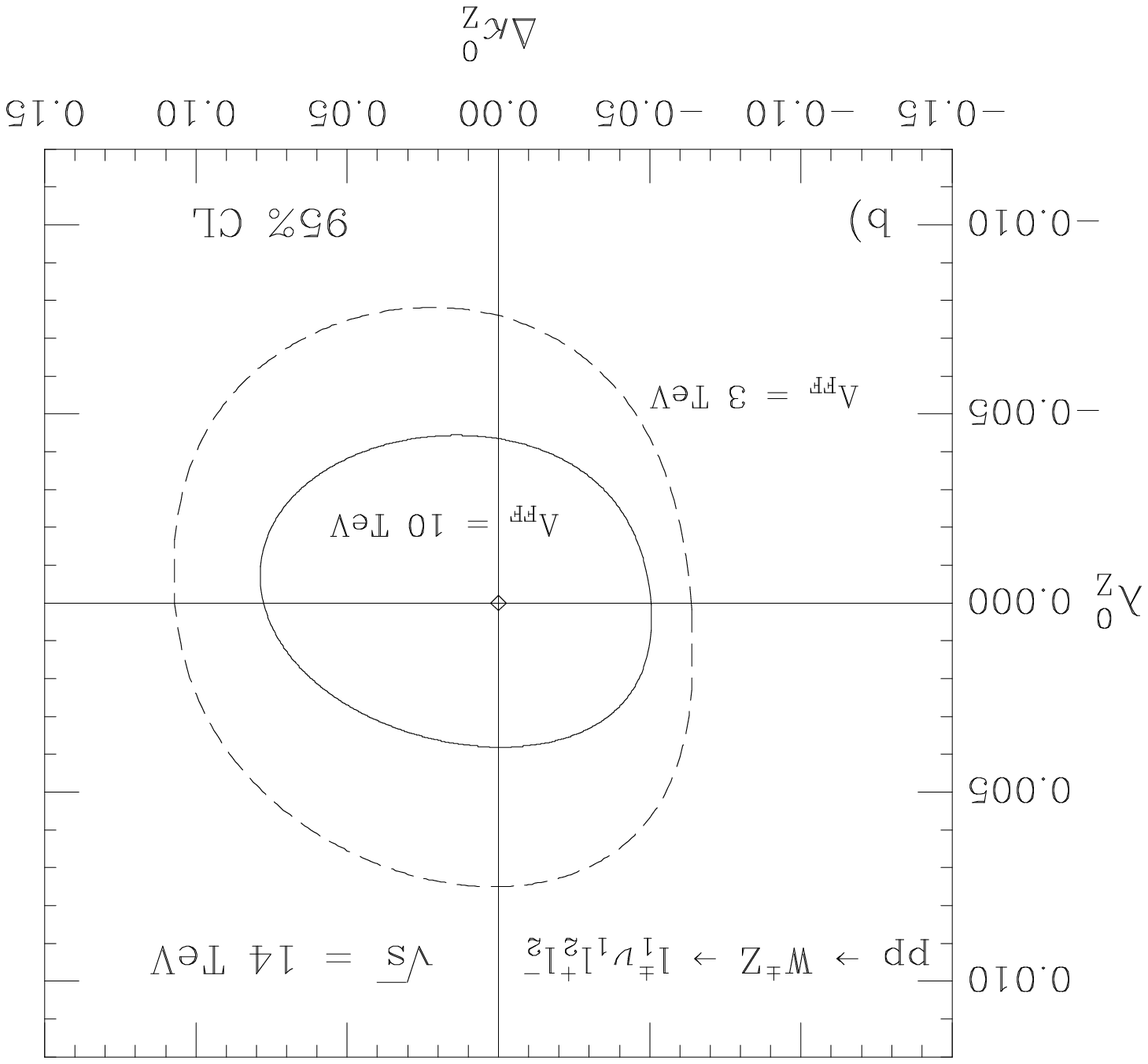}
\caption[]{95\% CL sensitivity limits from 
$W^\pm Z\to\ell_1^\pm\nu_1\ell_2^+\ell_2^-$ at the LHC $(a)$ in the 
HISZ scenario and 
$(b)$ if only $\Delta\kappa_Z$ and $\lambda_Z$ are allowed to deviate from
the Standard Model.}
\label{ellipseone}
\end{figure}

\begin{figure}[t]
\vskip 12cm
\includegraphics{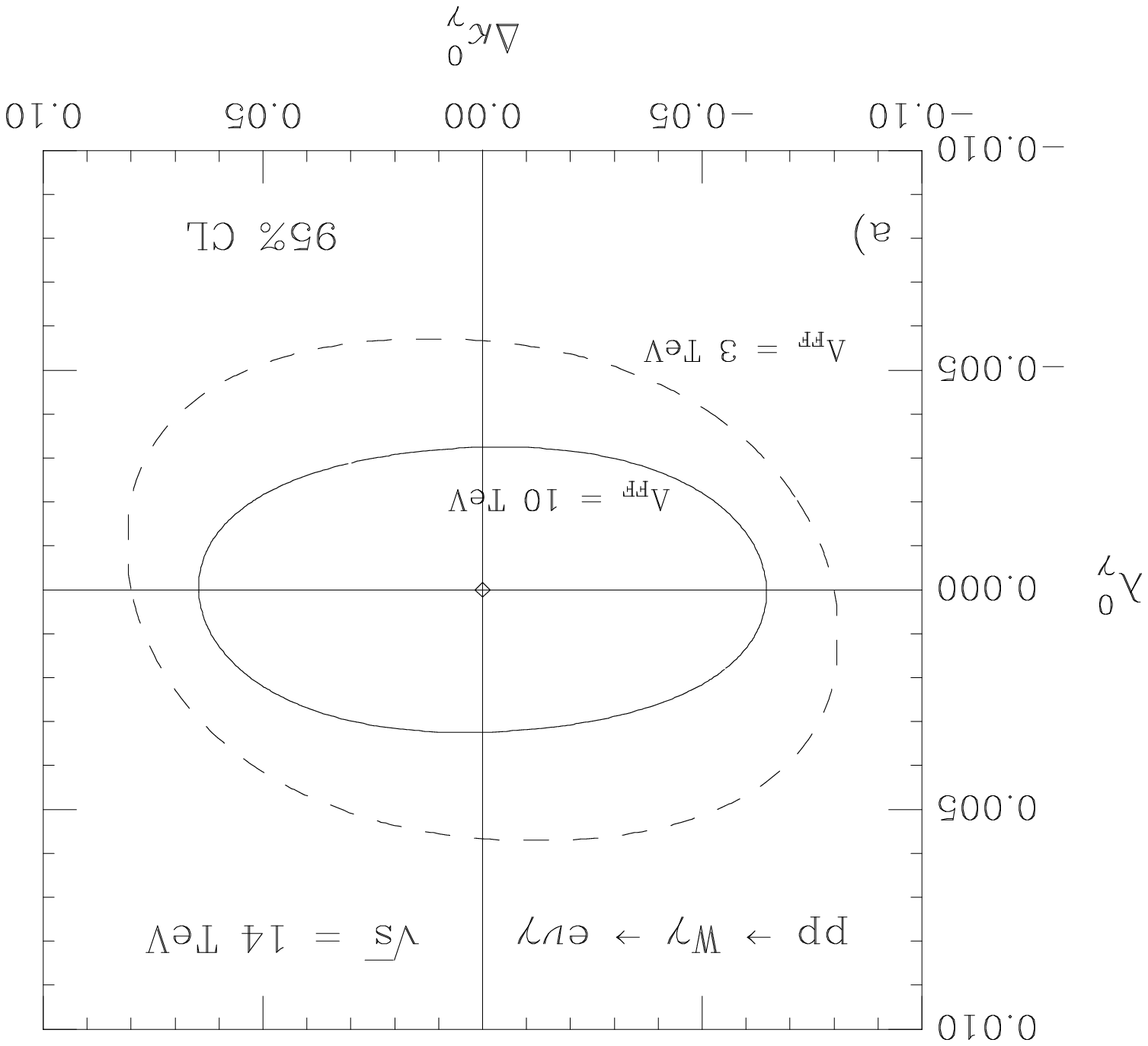}
\includegraphics{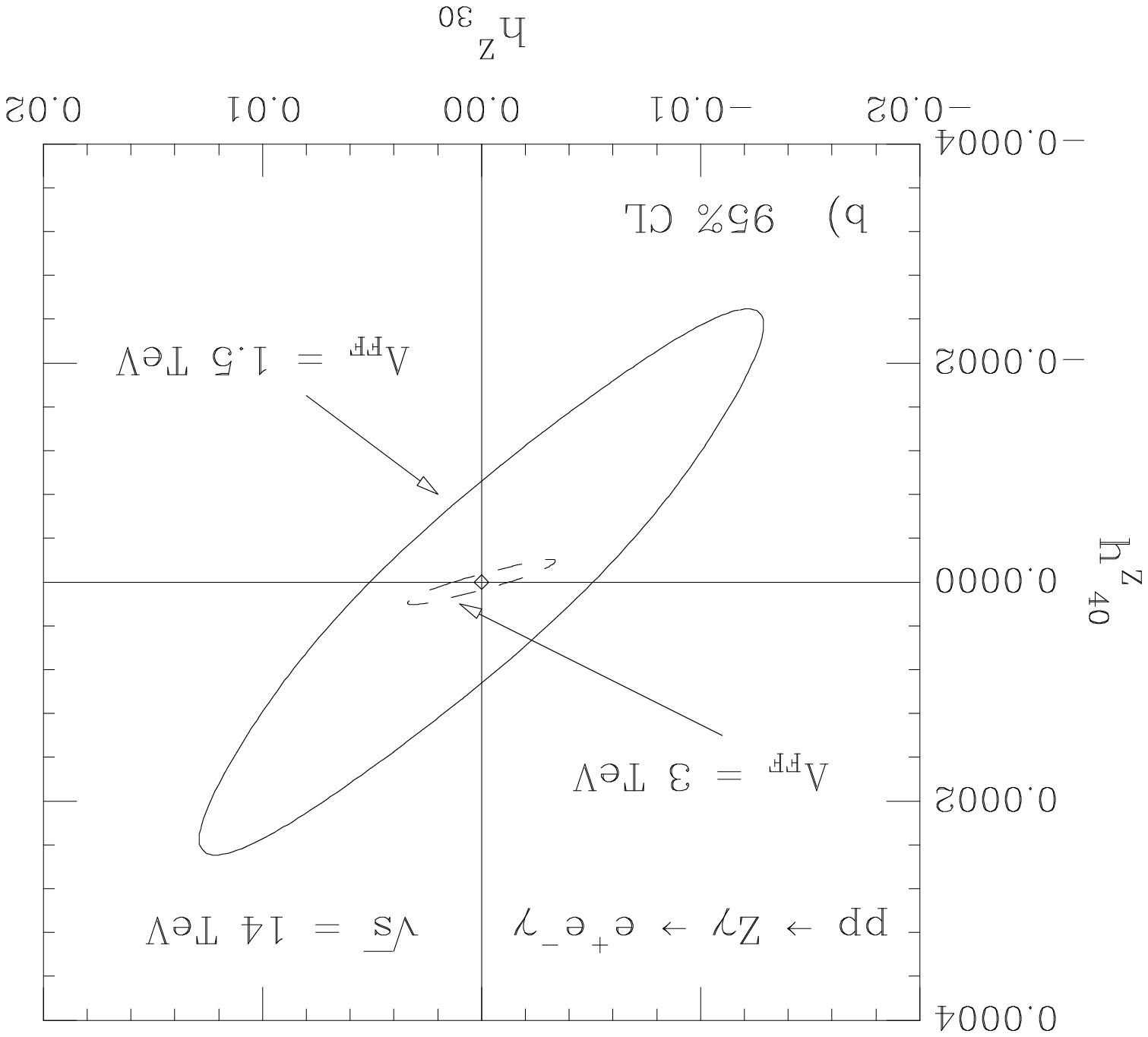}
\caption[]{95\% CL sensitivity limits for $(a)$ $WW\gamma$
couplings from $W\gamma$ production and $(b)$ $ZZ\gamma$ couplings from
$Z\gamma$ production at the LHC. Results are displayed for an integrated
luminosity of 100~fb$^{-1}$ and two different form factor scales.}
\label{ellipsetwo}
\end{figure}

Studies\cite{dpf} have also been carried out 
for the 1994 DPF Long Range Planning Workshop.
For $WZ$ states, the $eee\nu$ signal only was considered, and it was required
that $p_T^\ell > 25\,{\rm GeV/c}$, and $E_T^{miss} > 50\,{\rm GeV}$.  A binned
likelihood fit to the $p_T^Z$ distribution then yields limits on
$\Delta\kappa_Z$ and $\lambda_Z$ which are shown in Fig\ref{ellipseone}.
For $W\gamma$ and $Z\gamma$ states, a combination of ATLAS resolutions and CDF
efficiencies was assumed.  
It was required
that $p_T^\ell > 40\,{\rm GeV/c}$, 
$p_T^\gamma > 25\,{\rm GeV/c}$, $E_T^{miss} > 25\,{\rm GeV}$ ($W\gamma$ only),
and $m(\ell\ell\gamma) > 110\,{\rm GeV/c^2}$ ($Z\gamma$ only).
A separation of $\Delta R > 0.7$ between the lepton and photon was required, and
events with any jet with $E_T$ above $50\,$GeV were vetoed.  
A binned
likelihood fit to the $p_T^\gamma$ distributions then yields limits on
$\Delta\kappa_\gamma$, $\lambda_\gamma$, $h_3^Z$ and $h_4^Z$ which are 
shown in Fig\ref{ellipsetwo}.

\begin{table}[htb]
\begin{center}
\begin{tabular}{|c|c|c|} 
\hline
Channel&Study&Limit\\[1.mm] 
\hline\hline
$pp\to W^\pm\gamma\rightarrow e^\pm\nu\gamma$&DPF&$\Lambda_{FF}=3$~TeV:\\
&&$|\Delta\kappa_\gamma^0|<0.080$\\ 
&&$|\lambda_\gamma^0|<0.0057$\\ 
&&$\Lambda_{FF}=10$~TeV:\\ 
&&$|\Delta\kappa_\gamma^0|<0.065$\\
&&$\!\!|\lambda_\gamma^0|<0.0032$\\
&$\!$ATLAS$\!$&$\Lambda_{FF}=10$~TeV:\\ 
&&$|\Delta\kappa_\gamma^0|<0.04$\\
&&$|\lambda_\gamma^0|<0.0025$\\
\hline
$pp\to W^\pm Z\to\ell\nu\ell\ell$&DPF&$\Lambda_{FF}=3$~TeV:\\
$\ell=e,\mu$, HISZ \cite{hisz}&&
$\!\!\!-0.0060<\Delta\kappa_\gamma^0<0.0097\!\!$\\ 
&&$-0.0053<\lambda_\gamma^0<0.0067$\\
&&$\Lambda_{FF}=10$~TeV:\\ 
&&$\!\!\!-0.0043 < \Delta\kappa_\gamma^0<0.0086\!\!$\\ 
&&$-0.0043<\lambda_\gamma^0<0.0038$\\
\hline
$pp\to W^\pm Z\to\ell\nu\ell\ell$ &DPF&$\Lambda_{FF}=3$~TeV:\\
$\ell=e,\mu$, $\Delta g_1^Z=0$&&$-0.064<\Delta\kappa_Z^0<0.107$\\ 
&&$-0.0076<\lambda_Z^0<0.0075$\\
&&$\Lambda_{FF}=10$~TeV:\\ 
&&$-0.050 < \Delta\kappa_Z^0<0.078$\\ 
&&$-0.0043<\lambda_Z^0<0.0038$\\
&$\!$ATLAS$\!$&$\Lambda_{FF}=10$~TeV:\\
&&$|\Delta\kappa_Z^0|<0.07$\\
&&$|\lambda_Z^0|<0.005$\\
\hline
$pp\to Z\gamma\rightarrow e^+e^-\gamma$ & DPF &$\Lambda_{FF}=1.5$~TeV:\\
&&$|h^Z_{30}|<0.0051$\\
&&$|h^Z_{40}|<9.2\cdot10^{-5}$\\
&&$\Lambda_{FF}=3$~TeV:\\
&&$\!\!|h^Z_{30}|<0.0013$\\
&&$|h^Z_{40}|<6.8\cdot 10^{-6}$\\
\hline
\end{tabular}
\end{center}
\caption[]{Expected 95\% CL limits on anomalous $WWV$, $V=\gamma, \, Z$,
and $ZZ\gamma$ couplings from experiments at the LHC. Only one of the
independent couplings is assumed to deviate from the SM at a
time. The limits obtained for
$Z\gamma\gamma$ couplings almost coincide with those found for $h_3^Z$
and $h_4^Z$.}
\label{vvtable}
\end{table}

The limits obtained in all the above studies are summarized in
Table~\ref{vvtable}.  It will be possible to probe $WWV$ anomalous couplings 
with a precision of order $10^{-1}-10^{-3}$ if the form factor scale
$\Lambda_{FF} > 2\,{\rm TeV}$.  
This is sufficient to just reach
the interesting region where one may hope to see deviations
from the standard model given present limits on the scale of new physics.  

\section{Standard Model Physics}

\subsection{Top Quark Physics}

The potential for the study of the top quark at hadron colliders 
is already apparent.  Its recent discovery at the Tevatron
undoubtedly presages a long and fruitful program of top physics studies.
The LHC will be a top factory, with 
about 10$^7$ $t\overline t$ pairs produced per year at a luminosity
of $10^{33}\,{\rm cm}^{-2}{\rm s}^{-1}$.  This would result in about
200,000 reconstructed $t\overline t \to (\ell\nu b)(jjb)$ events and
20,000 clean $e\mu$ events.  

\subsubsection{Top Mass Measurement}

The top mass may be reconstructed from the $t\overline t \to (\ell\nu b)(jjb)$
final state using the invariant mass of the 3-jet system.  Problems arise
from the presence of backgrounds, from combinatorics, and from systematic
effects due to the detector and the theoretical models used.  
ATLAS\cite{atlas} have estimated that an 
accuracy of $\pm 3$~GeV could be attained.
By selecting very high-$p_T$ top quarks, where the decay products are boosted
and thus close, combinatorics may be reduced, and the mass measured to perhaps
$\pm 2$~GeV.  This measurement requires, of course, that the hadronic 
calorimetry be
calibrated to this level in the absolute energy scale and that its response be
stable over time.  CMS have investigated the possibility of in-situ calibration
of the jet response within top events by reconstruction of the hadronic
$W$ decays, a possibility already evident in the present CDF and D\O\ data.  

The mass may also be reconstructed from dilepton events.  ATLAS estimate that,
by selecting events with two leptons from $W$ decays and an additional lepton 
from $b$-decay, and plotting the invariant mass of the lepton pair originating
from the same top decay, the mass could be determined with a statistical
accuracy of $\pm 0.5$~GeV, and a total accuracy of about $\pm 2$~GeV. The
dominant systematic errors arise from uncertainties in the $b$-quark
fragmentation and are therefore complementary to the 3-jet system
which is dominated by calorimeter and jet systematics. 

\subsubsection{Search for Charged Higgs}

In extensions of the standard model with charged higgs bosons $H^\pm$, such as
in the MSSM, the decay $t \to b H^\pm$ may compete with the standard 
$t \to b W^\pm$ if kinematically allowed.  The $H^\pm$ decays to $\tau\nu$ or
$c \overline s$ depending on the value of $\tan\beta$.  Over 
most of the range $1 < \tan\beta < 50$, the decay mode
$H^\pm \to \tau\nu$ dominates.  The signal for $H^\pm$ production is thus an
excess of taus produced in $t\overline t$ events.

Both ATLAS\cite{atlascharg} and CMS\cite{cmscharg} have 
investigated the sensitivity to this excess.  
Top events with at least one isolated high-$p_T$ lepton are selected, and
the number having an additional tau compared with the number having an
additional $e$ or $\mu$.  Both studies used $b$-tagging to reduce the
backgrounds to top production.
Taus were identified
in a way very similar to that described earlier (in the section on
$A,H \to \tau \tau$
searches). 
The uncertainty in the tau excess is estimated to be $\pm 3$\%, dominated
by systematics. 
For an integrated luminosity of $10^4\,{\rm pb}^{-1}$, both ATLAS and CMS conclude that
over most of the $\tan\beta$ range, a signal can be observed at the $5\sigma$
level for $m_{H^\pm} < 130$~GeV, which corresponds to the region
$m_A \simle 120$~GeV in the $m_A,\tan\beta$ plane. 

\subsubsection{Rare Top Decays}

The large statistics available at LHC will provide sensitivity to other
non-standard or rare top decays.  As an example, ATLAS have investigated
the channel $t \to Zc$\cite{atlas}, which should occur at 
a negligible level in the SM. 
With an integrated luminosity of $10^5\,{\rm pb}^{-1}$, branching ratios
as small as $5\times 10^{-5}$ could be measured.

The TeV2000 study\cite{tev2k} estimates that LHC will 
attain a precision 2--3 times
better than TeV33 on the ratio of longitudinal to left-handed $W$'s produced in
$t$ decays.  This ratio is exactly predicted in the SM for a given top mass,
and is sensitive to non-standard couplings at the $t \to Wb$ vertex, such
as a possible $V+A$ contribution.  

\subsection{$b$ Physics}

The preceding sections have shown the importance of $b$-tagging in addressing
many of the high-$p_T$ physics goals of the LHC.  Both major detectors will
consequently have the capability to tag heavy flavor production through
displaced vertices.  This capability, together with the
$b$-quark production cross-section at the LHC, 
will enable them to also pursue a targeted but interesting program of
$b$-physics.  It can be assumed that CP violation in the $b-$quark system will have been observed before the LHC gives data.
Nevertheless the enormous rate will enable a very precise determination of $\sin 2\beta$ to be made using the decay $B_u\to
\psi K_S$. An error of $\pm 0.02$ can be expected after $10$ fb$^{-1}$ of integrated luminosity. It should also be possible to
measure $B_s\overline{B}_s$ mixing and to search for rare decays such as $B\to \mu\mu$.  

\section{Summary and Conclusions}

The LHC is unique among accelerators currently existing or under construction. It 
will have sufficient energy and luminosity to
enable vital discoveries to be made and will lead 
to insight into the mass generation mechanism of the standard model.
The very detailed simulation studies carried out 
by the ATLAS and CMS  collaborations enable one to make the following statements
with a high degree of confidence:-
\begin{itemize}
\item If the minimal standard model is correct and the 
higgs boson is not discovered at LEP II, it will be found at LHC.
\item If supersymmetry is relevant to the breaking of 
electroweak symmetry, it will be discovered at LHC and
many details of the particular supersymmetric 
model will be disentangled. 
\item If the Higgs sector is that of the minimal 
supersymmetric model, at least one Higgs decay channel
will be seen, no matter what the parameters turn out to be. 
In many cases, several Higgs bosons or decay channels will be seen.
\item If the electroweak symmetry breaking proceeds via some new strong interactions,
many resonances and new exotic particles will
almost certainly be observed. 
\item New gauge bosons with masses less than several TeV will be discovered or
ruled out. 
\end{itemize}

A great opportunity and a vast amount of excitement is promised to those 
physicists fortunate enough to be part of an LHC
experiment.

% The references should come at the very end of the paper. Delete or change
%   the fake bibitem below.
% The tags that are used with the bibitems will be used with the \cite{tag}
%   command in the text to automatically produce cross references.


\begin{thebibliography}{73}

\bibitem{atlas} 
Atlas Technical proposal, CERN/LHCC/94-43.

\bibitem{cms} 
CMS Technical proposal, CERN/LHCC/94-38.

\bibitem{ckm} M. Kobayashi, T. Maskawa, 
 {\it Prog. Theor. Phys.} 49:652 (1973).
N. Cabibbo, 

\bibitem{ktev}K. Arisaka, \etal, FERMILAB-FN-580 (1992).

\bibitem{bfactory} D. Boutigny, \etal, SLAC-R-95-457, M.T. Cheng, \etal, KEK Report BELLE-TDR-3-95.

\bibitem{standard-model} S. Glashow, \np{22}{579}{61};       S. Weinberg, \prl{19}{1264}{67};
      A. Salam, {\it in}: ``Elementary Particle Theory,'' W. Svartholm,
ed., Almquist and Wiksell, Stockholm (1968);
H.D. Politzer, \prl{30}{1346}{73});
D.J.  Gross and  F.E. Wilczek,  \prl{30}{1343}{73}).
\bibitem{dis} G. Miller, \etal, \pr{D5}{6528}{72}, A. Bodek, \etal, \prl{30}{1087}{73}.
\bibitem{bandc} J.J. Aubert, \etal, \prl{33}{1404}{74};
J.E. Augustin, \etal, \prl{33}{1406}{74}; G. Goldhaber, \etal, \prl{37}{255}{76}; S.W. Herb, \etal,
\prl{39}{252}{77}; D. Andrews, \etal, \prl{45}{219}{80}.
\bibitem{neutral-currents} F.J. Hasert, \etal, \pl{46B}{138}{73}.
\bibitem{3jets} R. Brandelik, \etal, \pl{86B}{243}{79}; D.P. Barber, \etal,  \prl{43}{830}{79};
C. Berger, \etal, \pl{86B}{418}{79};W. Bartel, \etal, \pl{91B}{142}{80}.
\bibitem{topdisc}F. Abe \etal \prl{73}{2667}{94},\pr{D52}{2605}{95}; 
S. Abachi \etal \prl{74}{2632}{95}.
\bibitem{wz}G. Arnison, \etal \pl{126B}{398}{83}, 
G. Arnison, \etal \pl{122B}{103}{83}.
%\bibitem{ztobbar}P.B. Renton, at the {\it XVII International
%Symposium on Lepton and Photon Interactions at High Energies}, Beijing,
%China, August 10-15, 1995. 
\bibitem{neutrino-troubles}For example see, L Wolfenstein, hep-ph/9604389.
\bibitem{cahn96} R.N. Cahn, R. N., LBL-38649 (1996), submitted to Rev. Mod. Phys.
\bibitem{Lee-quigg} C. Quigg, B.W. Lee and H. Thacker, \pr{D16}{1519}{77}
      M.~Veltman, {\it Acta Phys. Polon.} B8:475 (1977).
\bibitem{susy} For a review see, I. Hinchliffe, \ar{36}{505}{86}; 
\bibitem{technicolor} For a review see, K.D. Lane hep-9605257 (1996).
\bibitem{lhcbook} P. Lefevre, \etal, CERN/AC/95-05.
\bibitem{alice} ALICE Technical Proposal, CERN/LHCC/95-71.
\bibitem{lhcb} LHC-B Technical Proposal, CERN/LHCC/95-XX.
\bibitem{ehlq} E. Eichten, \etal,  \rmp{56}{579}{84}.
\bibitem{geant}CERN Program Library, GEANT.
\bibitem{higgslimit} Review of Particle Properties, \pr{D54}{1}{96}.
\bibitem{lepstudy}LEP report, CERN 96-01.
\bibitem{cmsgamgam} C.~Seez, CMS-TN/94-289 (1994).
\bibitem{atlasgamgam} D. Froidevaux. F. Gianotti and E. Richter Was,
ATLAS Note PHYS-NO-064 (1995),
F. Gianotti and I.~Vichon, ATLAS Note PHYS-NO-078 (1996).
\bibitem{atlaswh}L. Feyard and G Unal, EAGLE PHYS-NO- 01.
\bibitem{froid95} D. Froidevaux and E. Richer-Was, \zp{C67}{213}{95}
\bibitem{atlastag}S Haywood, ATLAS internal note INDET-NO-92 (1995), I Gravilenko, \etal
ATLAS internal note INDET-NO-115 (1995).
\bibitem{cdftag}F. Abe, \etal,  \pr{D51}{4623}{1995}.
\bibitem{cmszzstar}D.~Denegri, R.~Kinnunen and G.~Roullet,
CMS-TN/93-101 (1993).
\bibitem{atlaszzstar}J.-C. Chollet, \etal,  ATLAS internal 
note PHYS-NO-17 (1992), L. Poggioli, ATLAS Note PHYS-NO-066 (1995).
\bibitem{anderson}  Greg W. Anderson, Diego J. Castano \pr{D53}{2403}{96}.
\bibitem{bosman} M. Bosman and M. Nessi, ATLAS Note PHYS-NO-050 (1994).
\bibitem{cmsllnunu} N.Stepanov, CMS-TN/93-87 (1993); S.~Abdullin and 
N.~Stepanov, CMS-TN/94-179 (1994).
\bibitem{cmslnujj} S.~Abdullin and 
N.~Stepanov, CMS-TN/94-178 (1994).
\bibitem{baur} J. Ohnemus, \pr{D50}{1931}{94}.
\bibitem{zmushko} S. Zmushko, \etal,  ATLAS Note PHYS-NO-008 (1992).
\bibitem{MSSMmasses} M. Carena, M. Quiros, C.E.M. Wagner,
\np{B461}{407}{96}.
\bibitem{snowgamgam}S.~Abdullin, C.~Kao and N.~Stepanov, University of
Rochester UR-1475, July 1996.
\bibitem{cmstautau}R.~Kinnunen, J.~Tuominiemi, and D.~Denegri,
CMS-TN/93-98 (1993) and CMS-TN/93-103; C.~Seez, CMS-TN/93-84.
\bibitem{atlastautau}D. Cavalli, \etal,  ATLAS Note PHYS-NO-025 (1993).
\bibitem{froid96}E. Richer-Was, \etal,   ATLAS Note PHYS-NO-074 (1996).
\bibitem{cmsmumu}N.~Stepanov, CMS-TN/94-182 (1994).
\bibitem{sdc}R.M. Barnett and I. Hinchliffe LBL-28773 (1990).
\bibitem{cmssquark}M. Greiter, CMS-TN/94-319 (1994); I.Iashvili \etal,
CMS-TN/93-74 (1993).
\bibitem{cmssquarklep}L. Rurua and N.Stepanov, CMS-TN/94-203 (1994);
L. Rurua, CMS-TN/94-207 (1994).
\bibitem{isasusy}F. Paige and S. Protopopescu, ISAJET V5.04.
\bibitem{tevsusy}F. Abe, \etal \prl{75}{613}{95}, \prl{76}{2006}{96};
 S. Abachi \etal \prl{75}{618}{95}, \prl{75}{613}{1995}.
\bibitem{tev2k} `Report of the TeV2000 Study Group on Future Electroweak
Physics at the Tevatron,' D. Amidei and R.~Brock (eds.), Fermilab 1996. 
\bibitem{atlascal}A. Artamonov ATLAS internal note CALNO-065 (1994), 
\bibitem{paigeslept}H. Baer, \etal, \pr{D49}{3283}{94}.
\bibitem{susysnow} A. Bartl, \etal, To appear in Proceedings of 1996 DPF Summer study;
I. Hinchliffe \etal, hep-ph/9610544.
\bibitem{toypaige} F. Paige,  To appear in Proceedings of 1996 DPF Summer study.
\bibitem{jesper}J. Soderqvist, To appear in Proceedings of 1996 DPF Summer study.
\bibitem{tevwmass}F. Abe, \etal, \prl{75}{11}{95}, S. Abachi, \etal  FERMILAB-PUB-96-177-E.
\bibitem{weiming} W-M Yao, To appear in Proceedings of 1996 DPF Summer study.
\bibitem{paige}H. Baer, \etal, \pr{D52}{2746}{95}, \pr{D51}{1046}{95}.
\bibitem{paige-atlas}F. Paige, ATLAS Note PHYS-NO-085 (1996).
\bibitem{LEP} LEP Electroweak working group.
\bibitem{dine}M. Dine, A. Nelson and Y. Shirman, \pr{D51}{1362}{95}.
\bibitem{thomas}S Dimopoulos, \etal, SLAC-PUB- 96-7104.
\bibitem{chanowitz}M.S. Chanowitz and M.K. Gaillard, \np{B261}{379}{85}.
\bibitem{atlaswwnote} G. Azuelos, \etal, ATLAS Note PHYS-NO-033 (1994).
\bibitem{fouchez}D. Fouchez, ATLAS internal note PHYS-NO-160 (1994)
\bibitem{dpf}H. Aihara {\it et al., FERMILAB-Pub-95/031.}
\bibitem{hisz}K. Hagiwara, S. Ishihara, R. Szalapski, and D. Zeppenfeld,
               \pl{B283}{353}{92},\pr{D48}{2182}{92}.
\bibitem{smith96} J.R. Smith, CMS TN/95-179 (1995).
\bibitem{bess}R. Casalbouni, \etal,  {\it Int. J. Mod. Phys.}{\bf A4},1065 (1989).
\bibitem{tevzprime} F. Abe, \etal , \prl{74}{2900}{95}, \pr{D51}{949}{95}.
\bibitem{atlaszprime}A. Henriquez and L. Poggioli, ATLAS Note PHYS-NO-010 (1992).
\bibitem{atlascharg}D. Cavalli, \etal, ATLAS internal note PHYS-NO-053 (1994).
\bibitem{cmscharg}R.~Kinnunen, D. Denegri and J.~Tuominiemi,
CMS-TN/94-233 (1994).

\end{thebibliography}
\end{document}